\UseRawInputEncoding
\documentclass[journal]{IEEEtran}

\usepackage{amsmath,amssymb,amsfonts,amsthm,dsfont} 
\usepackage{graphics}
\usepackage[dvips]{graphicx}
\usepackage[table,usenames,dvipsnames]{xcolor}
\usepackage{subcaption}
\usepackage{tikz-cd}
\usepackage{algorithm2e}
\usepackage[breaklinks=true, colorlinks, bookmarks=true, citecolor=Black, urlcolor=Violet,linkcolor=Black]{hyperref}

\usepackage{hyperref}

\newtheorem{theorem}{Theorem}
\newtheorem{lemma}{Lemma}

\newtheorem{corollary}{Corollary}
\newtheorem{assum}{Assumption}
\newtheorem{assumption}{Assumption}
\newtheorem{example}[assum]{Example}
\newtheorem{definition}{Definition}
\newtheorem{remark}{Remark}

\newcommand{\ba}{\begin{align}}
\newcommand{\ea}{\end{align}}

\newcommand{\IR}{\mathbb{R}}
\newcommand{\eqdef}{\stackrel{\triangle}{=}}
\newcommand{\T}{\mbox{\scriptsize{\rm T}}}

\begin{document}

\allowdisplaybreaks

\title{Inverse Optimal Safety Filters} 

\author{Miroslav~Krstic
\thanks{M. Krstic is with the Department
of Mechanical and Aerospace Engineering, University of California at San Diego, La Jolla,
CA, 92093-0411 USA (e-mail: krstic@ucsd.edu).}}

\maketitle

\begin{abstract}
CBF-QP safety filters are pointwise minimizers of the control effort at a given state vector, i.e., myopically optimal at each time instant. But are they optimal over the entire infinite time horizon? What does it even mean for a controlled dynamic systems to be ``optimally safe'' as opposed to, conventionally ``optimally stable''? When disturbances, deterministic and stochastic, have unknown upper bounds, how should safety be defined to allow a graceful degradation under disturbances? Can safety filters be designed to guarantee such weaker safety properties as well as the optimality of safety over the infinite time horizon? We pose and answer these questions for general systems affine in control and disturbances and illustrate the answers using several examples. In the process, using the existing QP safety filters, as well as more general safety-ensuring feedbacks, we generate entire families of  safety filters which are optimal over the infinite horizon though they are conservative (favoring safety over `alertness') relative to the standard QP. 
\end{abstract}

%

\IEEEpeerreviewmaketitle

\section{Introduction}

\subsection{CBFs: a few highlights}

It was in two ways that the 2014 paper \cite{AmesCruiseControl}, along with its later journal version \cite{AmesAutomotive}, marked a watershed in the study of nonlinear control systems under state constraints. 

First, by advancing the notion of control barrier functions (CBF) proposed in \cite{Wieland}, it laid the foundation for a Lyapunov-like alternative to constraint-handling control design methods like classical optimal control, MPC \cite{rawlings2017model}, or barrier Lyapunov functions (BLF) \cite{10.1016/j.automatica.2008.11.017}. While similar in name to CBFs, BLFs represent a more conservative approach in which the system is actively repelled from the boundary, as opposed to being just slowed down in its approach to the boundary. Furthermore, neither MPC nor BLFs entail the notion of a nominal control, as the primary purpose for the application of the input, from which the constraint-handling design should deviate only when a safety violation is imminent. 

Second, \cite{AmesCruiseControl,AmesAutomotive} proposed, following inspiration from \cite{doi:10.1137/S0363012993258732}, that the conflict between safety and the said nominal control (equilibrium stabilization, trajectory tracking, or mere open-loop forcing of the system) be `mediated' using a quadratic program (QP), in which the deviation of the actual control input from the nominal input is penalized quadratically, while the linear inequality constraint comes from the linearity in control of the nonnegative sum of the derivative of the CBF with an appropriate decay margin that limits the rate of approach to the barrier. This approach to imparting safety on a controlled system, while also obeying the system operator's intent, has been the most influential legacy of \cite{AmesAutomotive,AmesCruiseControl}. Virtually all the work on CBF-based safety maintenance employs some form of QP-based redesigns of the nominal control, often referred to as ``safety filters.''

CBFs have since been used in a  range of  domains, including 
multi-agent robotics \cite{WangMagnusMultiRobot,SantilloMulti,MagnusCortesNonsmooth}, 
automotive systems \cite{JankovicDriverIntent,AmesCruiseControl,XuAmesLaneKeeping}, 
robust safety \cite{JANKOVIC2018359,AmesISSCBF,XuRobustness}, 
delay systems \cite{TamasCovid,DSCC,JankovicDelay,PrajnaMultiStateDelay}, 
and stochastic systems \cite{CLARK2021Stochastic,PrajnaStochastic,CooganStochastic}. 

Since CBFs define constraints and, as such, represent system outputs, when paired with system inputs they have relative degrees. For example, a position constraint, such as a relative distance between cars on the road, is of relative degree two in reference to an idealized accelerator input on a car
but of relative degree three or higher in reference to the actual engine throttle input. CBFs of high relative degree, under that name, were first studied  in the 2015 articles \cite{HsuBipedal,WuSreenathFirstCBFhighRelDeg} with  progress following in~\cite{nguyen2016exponential,XU2018195,xiao2019control,breeden2021high} and continuing. 
However, control designs for specific CBF of arbitrarily high relative degree already appear in the 2006 article \cite{krstic2006nonovershooting}, which  presents backstepping designs for regulation to the boundary of the safe set, referred to, at that time, as `non-overshooting control.'

\begin{table*}[t]	
\caption{Designs that expand the ``safety filter toolkit''  relative to the introductory CBF-QP \cite{AmesAutomotive}} 
\label{table-contributions}
\centering
\begin{tabular}{l|c|c|c|c|}
\cline{2-5}
& \multicolumn{2}{c|}{deterministic}
& \multicolumn{2}{c|}{stochastic}
\\ \cline{2-5}
& no disturbance & with disturbance & known noise covariance & unknown noise covariance
\\ \hline \hline
\multicolumn{1}{ |l| | }{QP and {\em not} inv. opt.} & \cite{AmesAutomotive} & Thm \ref{thm-QP-DSSf} &  Thm \ref{thm-basic-stoch-QP} &  Thm \ref{C3thm2.1s}
\\ \cline{1-5}
\multicolumn{1}{ |l||  }{QP and inv. opt.} & Cor \ref{C3thm8.1e} and \ref{C3thm8.1eB} & Thm \ref{C3thm8.1d} & Thm \ref{C3thm8.1es} &  
\\ \cline{1-5}
\multicolumn{1}{ |l||  }{{\em non-QP} and inv. opt.} & Thm \ref{C3thm8.1B} & Thm \ref{C3thm8.1} & Thm \ref{thm:inop} & Thm \ref{C5thm:inop}
\\ \cline{1-5}
\end{tabular}
\end{table*}

\subsection{$L_gh$ `safety filters'}

QP-based safety filters are reminiscent of the 1980's-era parameter projection used in adaptive control \cite[Appendix E]{krstic1995nonlinear}, which defines the safe set through a `zeroing CBF.' Between parameter projection and QP-based safety filters there are two differences and one key similarity. One difference is that, in parameter estimation, the plant  is simply a vector integrator (of the update law), as opposed to being a general nonlinear system affine in control. The other difference is that parameter projection is an extreme (discontinuous) form of a QP-based safety filter:  projection lets the nominal update  proceed unaltered up to the boundary of the safe set and then tangentially projects the update, allowing the trajectory to slide along the boundary if the nominal update directs the estimates outward. As for the key similarity between parameter projection and QP safety filters, projection also employs a CBF, as well as a quadratic program. As a result, it has an $L_gh$ factor, a hallmark of CBF-QP. More on this in Sec. \ref{sec-projection}. 

A factor of $L_gh$ is a tell-tale sign of potential optimality---not mere pointwise optimality, at a given point $x$ in the state space, but optimality over the infinite time horizon. The so-called ``$L_gV$ controllers'' have a storied history in nonlinear stabilization. Sontag's `universal formula' \cite{SONTAG1989117} was the first generally applicable $L_g V$ controller and is both pointwise and infinite-horizon optimal. Sepulchre, Jankovic, and Kokotovic \cite{sepulchre1997constructive}  produced a collection of results with such ``damping controllers'' and showed that every $L_gV$ controller---not just Sontag's formula---is optimal with respect to a meaningful cost functional if multiplied by a factor of two or more, which, in particular, indicates the controller's infinite gain margin. They also proved a nonlinear version of a $60^\circ$ phase margin: an $L_gV$ controller remains stabilizing when applied through any dynamical system of the form ${\rm Id} +{\cal P}$, where ${\rm Id}$ denotes identity and ${\cal P}$ denotes any strictly passive nonlinear system, which need not be input-to-state stable. 

Such  properties of $L_gV$ controllers inspired their further development under uncertainties. In \cite{661589}, for systems affine in control and disturbances, inverse optimal controllers were designed which solve a zero-sum game problem, in which the disturbance maximizes and the control minimizes a meaningful cost. In~\cite{935055,ITO200259}, global inverse optimality was augmented with local direct optimality.  In \cite{DENG1997151,940927} stochastic inverse optimal designs were introduced: $L_gV$ controllers for inverse optimal stabilization in probability in \cite{DENG1997151} and controllers that are inverse optimal for a zero-sum game relative to the unknown covariance acting as the opposing player in \cite{940927}. Finally, in \cite{LI19971459}, {\em adaptive} $L_gV$ controllers were designed which minimize a penalty not only on the plant's state and the input, but also on the parameter estimation error---thus far the only pairings of controllers and parameter estimators which  are not merely optimal `asymptotically'  but over the entire time horizon. In each of \cite{661589,DENG1997151,940927,LI19971459}, $L_gV$ controllers are designed not only for some classes of systems but for all suitably stabilizable systems, using Sontag-type formulae.

Given the $L_gh$ form of the CBF-QP safety filters \cite{AmesAutomotive,AmesCruiseControl}, it is imperative to ask the following questions. Are the CBF-QP safety filters inverse optimal? If not, can they be made optimal with respect to some meaningful cost functionals? What is meaningful to penalize when `mediating' safety and the execution of the user's nominal control design? 

To answer these questions, let us consult intuition. First, let us note that CLFs and CBFs are not the  opposites of each other: CLF is an energy-like, or norm-like function, whereas a CBF is a system output. However, for both CLFs and CBFs we are interested in their decays and growths. While the decay of a CLF  indicates convergence to an equilibrium, i.e., an improvement in desired performance, the increase of a CBF indicates movement away from the dangerous boundary of the safe set, i.e., an improvement in safety. Hence,  optimization should reward an increase in safety. 

Another hint  comes from terminology: if the $L_gV$ controllers got  nicknamed the ``damping controllers'' because they enhance the negativity of $\dot V$, the CBF-QP safety filters, which {\em reduce} the negativity of $\dot h$, should be nicknamed  ``anti-dampers''  among safety filters. In fact, the CBF-QP safety filters act precisely as pointwise worst-case disturbances, not unlike the optimal disturbances in $H_\infty$ control. 

In summary, optimal safety filters should be maximizing a reward function that is (1) proportional to the CBF and (2) negative definite in the deviation between the control applied and the nominal control.  In plain language, optimality should reward both safety and close adherence to the nominal control. 

Let us now return to the question---is CBF-QP inverse optimal? It is not. It is only optimal in a myopic sense, pointwise in $x$, but not over the infinite horizon. Infinite-horizon optimality has been pursued in \cite{9147721,9303896,9467052,almubarak2021hjb} but towards achieving optimal stabilization, not {\em optimal safety}.

Can we design safety filters that have a property of inverse optimality? Yes and, in the absence of a disturbance, such a redesign amounts to little more than 
multiplying the QP modification to the nominal control by a factor of two or higher. Plainly speaking, doubling the anti-damping of the CBF, i.e., doubling  safety, imparts inverse optimality.

\subsection{Nonlinear systems with disturbances: deterministic and stochastic}

Under deterministic disturbances, two main ideas have emerged. Robust CBFs \cite{JANKOVIC2018359} ensure safety under a disturbance with a known  bound. In input-to-State Safety (ISSf)  \cite{AmesISSCBF}, which mirrors input-to-state stability (ISS) \cite{28018}, the disturbance is bounded but potentially arbitrarily large and, being also unvanishing, may take the system outside of the safe set. Hence,  the CBF $h$ may assume negative values but in proportion to the size of the disturbance, with a class ${\cal K}$ gain from the disturbance to the `safety violation' $-h$. 

Controllers that render the safety violation $-h$  proportional to the disturbance are introduced in the 2006 work on non-overshooting control \cite{krstic2006nonovershooting} with a backstepping design for a high relative degree CBF. 

In the stochastic case, a general CBF-based safety analysis is presented in \cite{CLARK2021Stochastic}. A mean-non-overshooting tracking design for stochastic strict-feedback systems is given in \cite{WuquanStochasticNonovershooting}. 

In this paper we tackle four questions related to systems with disturbances. Two of the questions are the designs of QP-based safety filters for general nonlinear systems affine in deterministic or stochastic disturbances. The other two questions are the inverse optimal versions of safety filters under stochastic and deterministic disturbances. 

But what does inverse optimality mean in the presence of disturbances? Disturbances not  subject to a known upper bound dictate that optimality take a form of a two-player game, between the safety filter and the disturbance. While the safety filter's goal is to maintain safety, the goal of the disturbance is to erode safety, while investing as little of its energy as possible. This leads to cost functionals positive definite in the disturbance and proportional to the CBF, with the goal of the disturbance to minimize such a cost. 

{\bf\em Contribution Summary:} The expansion of the ``safety filter toolkit'' which we offer here, relative to the introductory CBF-QP \cite{AmesAutomotive}, 
is displayed in Table \ref{table-contributions}. 
We design safety filters which are deterministic and stochastic disturbance-based versions of the CBF-QP design. We also provide their modifications which ensure inverse optimality. Our safety filters are Nash equilibrium strategies, in balance with the Nash equilibrium strategies of the disturbances. 

\subsection{Safety framework}

We are unconcerned with stability in this paper. We consider a hierarchical scenario comprising (1) at the bottom layer, an ``operator'' ${\cal O}$, who only commands setpoints or open-loop reference signals; (2) at the middle layer, a designer ${\cal N}$ of a nominal feedback law $u_0$, which fulfills ${\cal O}$'s command in the absence of state constraints; and (3) at the top layer, a designer ${\cal S}$ of a safety filter $\bar u$ which ensures safety for a given nominal $u_0$. The barrier function $h(x)$ is known only to ${\cal S}$. 

The scenario commonly considered in CBF-CLF-QP  \cite{AmesCruiseControl,JANKOVIC2018359} has the setpoint  within the safe set. We allow ${\cal O}$ to, more generally, possibly command operation outside of the safe set (unknown to ${\cal O}$), and this makes the stability issue moot. As we shall see in Example \ref{ex-blowup}, even forward completeness may be in conflict with the nominal control and safety objectives. 

\subsection{Preview of inverse optimal safety filters: a scalar example} \label{sec-intro-example}

Consider the scalar system 
\begin{equation}
\dot x = u\,, \quad h(x) = - x\,. 
\end{equation}
Merely maintaining the positive invariance of $\{ x<0\}$ is achievable  with trivial  $u=0$ and even with  destabilizing  $u=x$. A good safety filter should keep $u $ close to the nominal  $u_0$ when $-x>0$ is comparatively large, i.e., when  $x$ is far from the  boundary $x=0$. 

The QP solution $\bar u_{\rm QP} = \min\{0, -u_0-x\}$ gives the safety filter $u= u_{\rm QP} = u_0 + \bar u_{\rm QP}  = \min\{u_0, -x\}$, which ensures safety with $\dot h \geq -h$ and is pointwise optimal in $x$ but is not optimal over the interval $0\leq t<\infty$. However, the modified QP safety filter $u=u^*_{\rm QP} = u_0+ {\bf 2}\bar u_{\rm QP} = u_0+2 \min\{0, -u_0-x\}= \min\{u_0, -u_0-2x\}$, namely,
\begin{equation}\label{eq-u*QP}
u = u^*_{\rm QP} = -x -|u_0+x|
\end{equation}
not only ensures safety but also, as we shall see in the general results later, {\em maximizes} the cost functional
$-x(+\infty)+\int_0^\infty \bigg(\min\{-x, u_0\} 
- \frac{(u - u_0)^2 }{ 4 \max\{0, u_0 + x\}}\bigg) \, {\rm d} t $,
and, equivalently, {\em minimizes} the cost functional
\begin{equation}\label{eq-invoptex-min}
x(+\infty) + \int_0^\infty \bigg(\max\{x, -u_0\} 
+ \frac{(u - u_0)^2 }{ 4 \max\{0, u_0 + x\}}\bigg) \, {\rm d} t \,,
\end{equation}
where we have simply suppressed the dependence on $t$ in $x(t),u(t),u_0(x(t),t)$ under the integrals for the sake of clarity.

The  functional 
\eqref{eq-invoptex-min} is meaningful. The term
$x(+\infty)$ is a ``terminal safety violation'' cost and the term $\max\{x, -u_0\} $ under the integral is a running safety violation cost. The term $(u - u_0)^2$ penalizes the deviation of $u$ from $u_0$ 
and its denominator inflicts an infinite penalty on  $u$ for possibly not remaining at the exactly the nominal $u_0$ when $u_0<-x$, which is when the nominal control is acting on its own to push the state away from the boundary $x=0$.  The value function of 
\eqref{eq-invoptex-min} 
is the `safety violation' $+x$,
which means that the optimizing safety filter results in the optimal cost
$+x_0<0$. 
In summary, 
\eqref{eq-invoptex-min} incentivizes both safety and ``alertness.''  

If the reader is unsettled by the non-smoothness of 
$\max$ in 
 \eqref{eq-invoptex-min}, or simply put off by the dogmatism of QP/min-norm control, an alternative inverse optimal safety filter is the Sontag formula-inspired $\bar u_{\rm S}= -\left(u_0+x + \sqrt{(u_0+x)^2+1}\right)$, which gives $u = u^*_{\rm S}= u_0+ \bar u_{\rm S}$, namely, 
\begin{equation}\label{eq-u*S}
u = u^*_{\rm S}= -x - \sqrt{(u_0+x)^2+1}\,,
\end{equation}
which {\em minimizes}
\begin{eqnarray}\label{eq-invoptex-minS}
x(+\infty) + \int_0^\infty \bigg[
{-u_0+x +\sqrt{(u_0+x)^2+1} \over 2}
 \qquad\qquad
\nonumber\\
\qquad\qquad+ {1\over 2}\frac{(u - u_0)^2 }{ u_0+x +\sqrt{(u_0+x)^2+1}}\bigg] \, {\rm d} t \,
\end{eqnarray}
and hence, like  
\eqref{eq-invoptex-min}, also maximizes safety and minimizes $u-u_0$. Note the similarity between the optimal QP filter \eqref{eq-u*QP} and the slightly more conservative optimal Sontag filter \eqref{eq-u*S}. Another variation on the Sontag formula is the safety filter $u= u_{\rm S}=u_0 + {1\over 2}\bar u_{\rm S} = {1\over 2} \left(u_0-x - \sqrt{(u_0+x)^2+1}\right)$, which is not inverse optimal but guarantees safety with $\dot h \geq - h$. 

\subsection{Contributions and organization}


Sections \ref{sec-DSSf}, \ref{sec-DSSf-CBF}, \ref{sec-DSSf-filter}, \ref{C3sec3}, and \ref{sec-DSSf-invopt} deal with deterministic disturbances. The main result on inverse optimality for safety filters is in Section \ref{C3sec3}. Safety filters of the special QP form, under disturbances, are presented in Section \ref{sec-DSSf-filter} and their inverse optimal versions are presented in Section \ref{sec-DSSf-invopt}.

For the reader less interested in the effects of disturbances (or overwhelmed by them), Sections \ref{sec-disturbance-free} and \ref{sec-disturbance-freeR} specialize the inverse optimality results to the disturbance-free case. It is in these sections that the points of the paper are most transparently evident. Section \ref{sec-disturbance-freeR} presents a Reciprocal CBF version of the QP safety filter that is endowed with inverse optimality. The Reciprocal CBF formulation of inverse optimal safety filter design is the closest to the traditional notion of optimal control---the task of control is minimization. 

Stochastic systems are dealt with in Sections \ref{sec-stochCBF}, \ref{sec-stochinvop}, and \ref{sec-stochinvopNS}. Stochastic safety filters, stochastic QP formulae, and inverse optimal achievement of safety under stochastic disturbances are all new notions in the literature. Additionally, the notion of stochastic safety under non-unity covariance, where covariance is time dependent and of unknown bound, which is the subject of Section \ref{sec-stochinvopNS}, is a new topic in the safety literature. Sections \ref{sec-stochinvop} and \ref{sec-stochinvopNS} are written in a contrasting fashion: the former dealing with the easier but still novel case of inverse optimal safety filters for unity-intensity stochastic disturbances, and the latter dealing with the same topics but with covariance whose intensity is arbitrary and incorporated in the cost functional---rewarded for making the system less safe but penalized when its energy is large. The inverse optimality results in Sections \ref{sec-stochinvop} and \ref{sec-stochinvopNS} are given in the traditional mean sense, as in conventional stochastic optimal control. 

In Section \ref{sec-projection} we consider parameter estimation and contrast the classical QP-based projection operator with a novel safety filter which, unlike projection, is continuous and also inverse optimal. The value of the section is in its treatment in a particularly simple control problem---a vector integrator---and in illuminating more clearly than possible on any more complex problem the notion of inverse optimal safety filters, while also shedding light on the classical projection, in the sense of what it actually lacks (continuity and optimality). 

In Section \ref{sec-boundary} we return to non-overshooting control \cite{krstic2006nonovershooting}, i.e., regulation to the safety boundary, with a twofold interest. First, clarifying the relation between non-overshooting controllers and safety filters. Second, providing a major generalization of non-overshooting control---beyond the strict-feedback systems in \cite{krstic2006nonovershooting} and in the presence of deterministic or stochastic disturbances \cite{WuquanStochasticNonovershooting}. 

\paragraph*{Notation.} Let $a<0<b$. A continuous function $\gamma: (a,b) \rightarrow\IR$  with $\gamma(0)=0$ is of extended class ${\cal K}_{(a,b)}$ if it is strictly increasing. A continuous function $\beta: (a,b) \times \IR_{\geq 0} \rightarrow\IR$ is of class ${\cal KL}_{(a,b)}$ if it is of class ${\cal K}_{(a,b)}$ in its first argument and has a zero limit as its second argument goes to infinity. 

\section{Disturbance-to-State Safety} \label{sec-DSSf}

We start with definitions of a barrier function and  safe set. 

\begin{definition}\label{def-CBF}
The scalar-valued differentiable function $h: \IR^n \rightarrow \IR$ with $\inf_{x\in\IR^n} h(x)<0$ and $\sup_{x\in\IR^n} h(x) >0$ is referred to as a {\em barrier function candidate}. The set ${\cal C} =\{\left. x\in\IR^n \ \right| \ h(x)\geq 0\}$ is referred to as a {\em safe set}.  
\end{definition}

\begin{assumption}\label{ass-setC}
${\cal C}$ is 
without isolated points. 
\end{assumption}

Consider now the disturbance-driven system
\begin{equation} \label{C3eq2.4a}
\dot x = f (x) + g_1 (x)d\,, \quad d\in\IR^{m_1}\,.
\end{equation}

\begin{definition}\label{def-DSSf}
The set ${\cal C}$ of the system (\ref{C3eq2.4a}) is said to be {\em disturbance-to-state safe} (DSSf) if 
\begin{equation} \label{C3eq2.5b}
h(x(t)) \geq \beta(h(x_0),t) - \rho\left(\sup_{0\leq \tau\leq t} |d(\tau)|\right)\,, \ \ \forall t\geq 0\,.
\end{equation}
where the function $\rho\in {\cal K}$ is referred to as the {\em DSSf gain function} and $\beta
\in 
{\cal KL}_{(\inf h(\xi),\sup h(\xi))}=:{\cal KL}_h$. 
\end{definition}

This property is not new. Controller design ensuring DSSf, using backstepping for non-overshooting control, goes as far back as  2006 in the paper \cite{krstic2006nonovershooting}---see the safety bound (61) of Theorem 3 with a disturbance of unlimited unknown bound $\bar d$, as well as the safety bound (90) of Proposition 1 with an observer-based non-overshooting controller. 

The property 
\eqref{C3eq2.5b}, with $\sup$ over infinite time, is referred to as input-to-state safety (ISSf) by \cite{AmesISSCBF,lyu2020smallgain}. Since the property 
relates to the disturbance input, not a control input, we prefer the term DSSf over ISSf. Robust nonlinear control  has been plagued by terminological confusion relating to ``{\em input}-to-state stability,'' ISS, where the term ``input'' is used for what is actually a {\em disturbance}. We do not wish the safety literature to have to endure a similar conflation of disturbance and control. 

The following 
slightly modified \cite[Definition 4]{lyu2020smallgain}.

\begin{definition} The function $h$ is called a {\em DSSf barrier function} (DSSf-BF) if there exist a function $\rho: [0,+\infty) \rightarrow [0, -\inf h(\xi))$ of class ${\cal K}$ and a function $\alpha$ in   
${\cal K}_{(\inf h(\xi),\sup h(\xi))}$ 
such that, for all $x\in \IR^n, d\in \IR^{m_1}$, 
\begin{equation} \label{C3eq2.5a}
\begin{array}{c}
 \min\left\{ 0, h(x)\right\} \leq  - \rho(|d|) \\[0.3cm]
 \Rightarrow\quad L_f h+L_{g_1}h d \geq -  \alpha(h)\,. 
\end{array}
\end{equation}
\end{definition}

The following result is a  variation on \cite[Theorem 1]{lyu2020smallgain}, proved by adapting  \cite[Theorem 2.2]{Krstic-Deng-book} and \cite[Theorem 1]{AmesISSCBF}. 

\begin{lemma}\label{lem:DSSf}
For the system (\ref{C3eq2.4a}), if there exists a DSSf-BF $h$ such that \eqref{C3eq2.5a} holds, then the system is DSSf with $\beta(r,t)$ in \eqref{C3eq2.5b} defined by the solution to $\dot{\underline h} = -\alpha(\underline h),\ \underline h(0) = r$.
\end{lemma}

For converse barrier certificates, see \cite{8814799}. 

\section{DSSf-CBFs} \label{sec-DSSf-CBF}

Consider now, with loc. Lipschitz  $f,g_1,g_2$, the system
\begin{equation} \label{C3eq2.4}
\begin{array}{rcl}
\dot x & = & f (x) + g_1 (x)d + g_2 (x) u\,, \quad u\in \IR^{m_2}\,,
\end{array}
\end{equation}

\begin{definition}	\label{C3def2.1}
A scalar differentiable function $h$ is called a {\em DSSf-control barrier 
function (DSSf-CBF)}\ for (\ref{C3eq2.4}) if there exists a class ${\cal K}$ function $\rho: \IR_{\geq 0} 
\rightarrow [0, -\inf h(\xi))$  and $\alpha\in {\cal K}_{(\inf h(\xi),\sup h(\xi))}=:{\cal K}_h$ such that, 
for all $x\in \IR^n, d\in \IR^{m_1}$, 
\begin{equation} \label{C3eq2.5}
\begin{array}{c}
 \min\left\{ 0, h(x)\right\} \leq  - \rho(|d|) \\[0.3cm]
 \Rightarrow\ \ \displaystyle{\sup_{u\in\IR^{m_2}}\left\{L_f h+L_{g_1}h d + L_{g_2}h u \right\}}  \geq  - \alpha(h). 
\end{array}
\end{equation}
\end{definition}

The following  is obtained by adapting  \cite[Lemma 2.1]{661589}.

\begin{lemma}\label{C3lem2.1}
A pair $(h, \rho)$ satisfies (\ref{C3eq2.5}) 
if and only if
\begin{align}
 L_{g_2} h(x) = 0 \hspace{0.4cm} \Rightarrow \quad \omega(x)\geq 0\label{C3eq2.6}
\end{align}
where
\begin{equation}\label{C3eq8.19}
\omega(x) = L_f h - \left| L_{g_1}h\right| \rho^{-1}(\max\{0, -h(x)\})+\alpha(h(x))\,.
\end{equation}
\end{lemma}

DSSf-CBFs, which do not require the disturbance to be in a known compact set, are different than Robust CBFs \cite{JANKOVIC2018359,choi2021robust}. That is the very purpose of the antecedent in the implication \eqref{C3eq2.5} and the term $ \rho^{-1}(\max\{0, -h(x)\})$ in \eqref{C3eq8.19}.



\begin{theorem}\label{C3thm2.1}
If there exists a DSSf-CBF, the system (\ref{C3eq2.4}) is rendered DSSf using the following Sontag-type control law:\footnote{See also the proof of Therem 3.2 in \cite{661589} and Remark 5 of \cite{AmesISSCBF}.}
\begin{eqnarray}\label{C3eq8.18}
u= u_{\rm S}(x) = (L_{g_2}h)^{\T}\left\{ \begin{array}{ll}
\displaystyle{
\kappa(x)}, 
& (L_{g_2}h)^{T} \neq 0 \vspace{3ex}\\
0, & (L_{g_2}h)^{\T} = 0\,,
\end{array}\right.
\end{eqnarray}
where, with $\omega(x)$ defined in \eqref{C3eq8.19},
\begin{eqnarray}\label{C3eq8.18kappa}
\kappa(x) 
&=& \frac
{-\omega + \sqrt{\omega^2+\left(L_{g_2}h(L_{g_2}h)^{\T}\right)^2}}
{L_{g_2}h (L_{g_2}h)^{\T}} 
\nonumber\\
&=& \frac
{L_{g_2}h (L_{g_2}h)^{\T}} 
{\omega + \sqrt{\omega^2+\left(L_{g_2}h(L_{g_2}h)^{\T}\right)^2}}\,.
\end{eqnarray}
\end{theorem}

\noindent{\bf Proof.} 
We substitute (\ref{C3eq8.18}) into (\ref{C3eq2.4}) and get
\begin{align}
\dot h =\ & L_fh + L_{g_1}h d -\omega +\sqrt{\omega^2
+\left(L_{g_2}h(L_{g_2}h)^{\T}\right)^2} \nonumber\\
\geq\ & -\alpha(h(x))
+|L_{g_1}h|\left[\rho^{-1}(\max\{0, -h(x)\}) - |d|\right]\,.
\end{align}
For $\min\left\{ 0, h(x)\right\}\leq -\rho(|d|)$ we thus have
\begin{equation}
\dot h = L_{f+g_2\alpha_{\rm S}} + L_{g_1} h d
\geq -\alpha(h(x)) \,,
\end{equation}
which, thanks to Lemma~\ref{lem:DSSf}, completes the proof of DSSf.
\hfill $\Box$

\section{Disturbance-to-State Safety Filter}\label{sec-DSSf-filter}

Now we turn our attention to 
the simultaneous objectives of maintaining safety 
and  deviating as little as possible from the nominal $u_0(x,t)$. For that purpose we rewrite \eqref{C3eq2.4} as
\begin{equation} \label{C3eq2.4c}
\dot x = f (x) +g_2 (x) u_0+ g_1 (x)d + g_2 (x) (u-u_0)\,. 
\end{equation}
Let a DSSf-CBF $h(x)$ be available, with associated $(\rho,\alpha)$. Accounting for the inclusion of $u_0$ into the drift vector field  \eqref{C3eq2.4c}, we modify \eqref{C3eq8.19} as 
\begin{eqnarray}\label{C3eq8.19nom}
\omega(x,u_0) &=& L_{f+g_2 u_0}h  - \left| L_{g_1}h\right| \rho^{-1}(\max\{0, -h(x)\})
\nonumber\\
&& +\alpha(h(x))\,.
\end{eqnarray}
Then, we introduce the QP problem
\begin{eqnarray}\label{eq-QPquad}
&\bar u_{\rm QP} = \arg\min_{v\in\IR^{m_2}} |v|^2 \quad \mbox{subject to} &\\[2mm]
&\omega(x,u_0) + L_{g_2} h v \geq 0\,.&  \label{eq-QPineq}
\end{eqnarray}
The well-known explicit solution to this problem is \cite{doi:10.1137/S0363012993258732}
\begin{equation}\label{eq-QP-DSSf}
\bar u_{\rm QP} = \left\{ \begin{array}{ll}
0, 
&  \omega(x,u_0)\geq 0 \vspace{3ex}\\
-\displaystyle{\omega(x,u_0)\over |L_{g_2}h|^2} (L_{g_2}h)^T, &\omega(x,u_0) <0\,.
\end{array}\right.
\end{equation}

\begin{remark}  \label{rem-Lg2h=0}\em 
Regarding the possible division by $L_{g_2}h=0$ in the second case of \eqref{eq-QP-DSSf}, we recall that, by Lemma \ref{C3lem2.1}, every DSSf-CBF satisfies the implication $L_{g_2} h = 0 \Rightarrow \omega(x,u_0) \geq 0$, which is equivalent to the implication $\omega(x,u_0) < 0 \Rightarrow L_{g_2}h \neq 0$, and this precludes $L_{g_2}h$ being zero in the second case of \eqref{eq-QP-DSSf}, i.e., a division by zero is not possible. 
\end{remark}

While bounded, \eqref{eq-QP-DSSf} is not necessarily continuous at points where $L_{g_2}h(x)=0$. 
When the nominal $u_0$ is only a function of $x$, continuity can be ensured  by assuming the following.

\begin{assumption}  \label{rem-Lg2h-scp} 
For a given locally Lipschitz $u_0:\IR^n\rightarrow \IR^{m_2}$,  system \eqref{C3eq2.4c} satisfies the {\em small control property} (SCP) \cite{SONTAG1989117,sepulchre1997constructive,661589,JANKOVIC2018359}, i.e., there exists a (not necessarily known) continuous  $\bar u_c(x)$ such that $L_{g_2}h(x)=0 \Rightarrow \bar u_c(x)=0$ and 
\begin{eqnarray}\label{C3eq8.19nom-scp}
\omega_c(x) &=& L_{f+g_2 (u_0+\bar u_c)}h  - \left| L_{g_1}h\right| \rho^{-1}(\max\{0, -h(x)\})
\nonumber\\
&& +\alpha(h(x))\geq 0\,.
\end{eqnarray}
\end{assumption}

From \eqref{C3eq8.19nom-scp} it follows, for $\omega$ defined in \eqref{C3eq8.19nom}, that $\omega\leq 0 \Rightarrow |\omega|\leq |L_{g_2}h | |\bar u_c|$, from which the continuity follows for \eqref{eq-QP-DSSf}, as well as for the Sontag  controller in Theorem \ref{C3thm2.1} (and \ref{C3thm2.1s}). With the SCP, \eqref{eq-QP-DSSf} and \eqref{C3eq8.18} are also locally Lipschitz on the open set $L_{g_2} h(x) \neq 0$. 

With the QP safety filter \eqref{eq-QP-DSSf}, we have the following result. 

\begin{theorem}\label{thm-QP-DSSf}
The control law
\begin{equation}\label{eq-u0+QP}
u = u_0 + \bar u_{\rm QP}(x,u_0)
\end{equation}
with $\bar u_{\rm QP}(x,u_0)$ defined in \eqref{eq-QP-DSSf} and $\omega(x,u_0)$ defined in \eqref{C3eq8.19nom} renders the system \eqref{C3eq2.4c} DSSf with respect to the DSSf-CBF $h(x)$, with a gain function $\rho$, i.e., there exists $\beta\in{\cal KL}_h$ such that, for all $t\geq 0$,
\begin{equation} \label{C3eq2.5bQP}
h(x(t)) \geq \beta(h(x_0),t) - \rho\left(\sup_{0\leq \tau\leq t} |d(\tau)|\right)\,.
\end{equation}
\end{theorem}

\noindent{\bf Proof.} 
We substitute (\ref{eq-u0+QP}) and (\ref{eq-QP-DSSf}) into (\ref{C3eq2.4}),  get
\begin{align}
\dot h  =\ &  L_{f+g_2 u_0}h + L_{g_1} h d + L_{g_2} h \bar u_{\rm QP}
\nonumber\\
=\ &  -\alpha(h(x)) +\omega +\max\left\{0, -\omega\right\}
\nonumber\\
\ &  +|L_{g_1}h|\rho^{-1}(\max\{0, -h(x)\}) +L_{g_1} h d 
\nonumber\\
\geq\ & -\alpha(h(x)) +\max\left\{\omega,0\right\}
\nonumber\\ \ &
+|L_{g_1}h|\left[\rho^{-1}(\max\{0, -h(x)\}) - |d|\right]\nonumber\\
\geq\ & -\alpha(h(x)) 
+|L_{g_1}h|\left[\rho^{-1}(\max\{0, -h(x)\}) - |d|\right]\,,
\end{align}
and invoke Lemma \ref{lem:DSSf}. 
\hfill $\Box$

\begin{example}\label{ex-1+x2d}\em
Consider the system 
\begin{eqnarray}
\dot{x} & = & u+(1+x^2) d \label{C3exa8.1a}
\end{eqnarray}
with a DSSf-CBF $h(x) = - x$. For some $\rho\in{\cal K}_\infty$, \eqref{C3eq8.19nom} is $\omega = - u_0  - \rho^{-1}(\max\{0, x\}) +\alpha(h(x))$ and 
the QP formula \eqref{eq-QP-DSSf} gives $\bar u_{\rm QP} = \min\left\{ 0, -u_0 -\rho^{-1}(\max\{0, x\})+\alpha(h(x))\right\}$. Taking, e.g., $\alpha(h) = h$, the overall feedback \eqref{eq-u0+QP}, given by $u = \min\left\{ u_0 , -\rho^{-1}(\max\{0, x\})-x\right\}$, guarantees, $\forall\rho\in{\cal K}_\infty$, 
\begin{equation}
x(t) \leq {\rm e}^{-t}x_0 +\rho\left(\sup_{0\leq \tau\leq t} |d(\tau)|\right)\,, \quad \forall t\geq 0\,.
\end{equation}
\hfill $\Box$
\end{example}

\begin{example}\label{ex-blowup}\em
The safety filter \eqref{eq-u0+QP} ensures safety but, on its own, does not guarantee forward completeness. For instance, the example in Section \ref{sec-intro-example}, where $\dot x = u, h(x) = - x$, results in the QP safety filter $u= \min\{u_0, -x\}$. If the nominal control happens to be $u_0 = x^3$, the resulting overall control is $u= \min\{x^3, -x\}$. Within the safe set $x<0$ this feedback becomes simply $u = x^3$ and gives a closed-loop system $\dot x = x^3$, which has a finite escape time. While this might at first disappoint, it should not. The safety filter ensures both safety and the exact conformity with the nominal control $u_0 = x^3$. If the user wishes to drive the solution to $-\infty$ in finite time, this is what the user gets with this safety filter. 
\hfill $\Box$
\end{example}

Hence, insisting on forward completeness may contradict the nominal objective and is not implied by safety. Nevertheless, for reasons of being able to state results like \eqref{C3eq2.5bQP} for all $t\geq 0$, we seek conditions that ensure forward completeness. One way is to assume {\em unboundedness observability}  \cite{ANGELI1999209}. 

\begin{assumption}  \label{rem-Lg2h-uo} 
For system \eqref{C3eq2.4} with BF $h$ and nominal  $u_0$ there exists a proper, smooth  $U:\IR^n\rightarrow \IR_{\geq 0}$  
such that 
\begin{eqnarray}\label{C3eq8.19nom-uo}
L_{f+g_1 d + g_2 u}U &\leq& U +\sigma_1\big(\max\{0, -h(x)\} + |d|\big)
\nonumber\\ && +\sigma_2(|u- u_0|) 
\end{eqnarray}
for some $\sigma_1,\sigma_2\in {\cal K}_\infty$  and for all $x\in\IR^n, d\in\IR^{m_1}, u\in\IR^{m_2}$. In addition,  for some $M>0$, the feedback $u_c$ in Assumption \ref{rem-Lg2h-scp}  satisfies $|\bar u_c(x)|\leq \sigma_2^{-1} (M U(x)) ,\ \forall x\in\IR^n$.
\end{assumption}

With this assumption we ensure that, once we prove safety, namely, that $h(x(t))\geq - \rho\left(\sup_{t\geq 0} |d(t)|\right)$ holds, it follows that $\max\{0, -h(x(t)) \leq \rho\left(\sup_{t\geq 0} |d(t)|\right)$ and, from \eqref{C3eq8.19nom-uo} and \eqref{eq-QP-DSSf}, that $\dot U \leq (1+M)U + \sigma_1\big(\rho\left(\sup_{t\geq 0} |d(t)|\right) + \sup_{t\geq 0} |d(t)|\big)$, which implies forward completeness. The condition $|\bar u_c(x)|\leq \sigma_2^{-1} (U)$ in Assumption \ref{rem-Lg2h-uo}  is undoubtedly strong but the alternative routes to ensuring the existence of solutions are even less appealing.

Due to limited space, in the rest of the paper we do not belabor regularity and existence issues. Assumptions like Assumptions \ref{rem-Lg2h-scp} and \ref{rem-Lg2h-uo} can ensure these properties for all our safety filter designs, but at the expense of  restricting $u_0$. 

\begin{remark}  \label{rem-Lg2h=}\em 
A more compact way to write \eqref{eq-QP-DSSf} is
\begin{equation}\label{eq-QP-DSSfsf}
\bar u_{\rm QP} = (L_{g_2}h)^T
\displaystyle{\max\left\{0, -\omega(x,u_0)\right\}\over |L_{g_2}h|^2} \,,
\end{equation}
however, this expression is not equivalent to \eqref{eq-QP-DSSf} when $L_{g_2}h=0$ as it is not apparent in \eqref{eq-QP-DSSfsf} that there cannot be a division by $L_{g_2}h=0$ when $\omega <0$. Nevertheless, since controls like \eqref{eq-QP-DSSf} appear in our paper at least half a dozen times, for the sake of compactness we write \eqref{eq-QP-DSSf} as \eqref{eq-QP-DSSfsf}, counting on the reader to recall that this expression is, in fact, zero when $L_{g_2}h=0$.
\end{remark}

For the reader's future convenience, we point out that an alternative representation of 
\eqref{eq-u0+QP} with the safety filter \eqref{eq-QP-DSSf} is
\begin{equation}\label{eq-u0+QP-DSSf}
u = \left\{ \begin{array}{ll}
u_0, 
&  \omega(x,u_0)\geq 0 \vspace{3ex}\\
\chi_0(x) u_0 + \chi_1(x) , &\omega(x,u_0) <0\,
\end{array}\right.
\end{equation}
where
\begin{align}
\chi_0(x) &= I - \displaystyle{(L_{g_2}h)^T L_{g_2}h \over |L_{g_2}h|^2} 
\\
\chi_1(x) &=-(L_{g_2}h)^T\displaystyle{\omega_1(x) \over |L_{g_2}h|^2} 
\\ 
\omega_1(x)&=L_{f} h - \left| L_{g_1}h\right| \rho^{-1}(\max\{0, -h(x)\})+\alpha(h(x)) \,.
\end{align}

The `half-Sontag' formula  also generates min-norm control. 

\begin{theorem}
The feedback
\begin{equation}\label{eq-1/2S}
u = u_0 + {1\over 2} u_{\rm S}\,,
\end{equation}
with $u_{\rm S}$ defined in \eqref{C3eq8.18}, \eqref{C3eq8.18kappa} and $\omega$ defined in 
\eqref{C3eq8.19nom} renders the system \eqref{C3eq2.4c} DSSf and is the pointwise minimizer of $|v|^2$ subject to the following constraint more conservative than \eqref{eq-QPineq}:
\begin{eqnarray}
{1\over 2} \left(\omega -\sqrt{\omega^2+\left(L_{g_2}h(L_{g_2}h)^{\T}\right)^2}\right) + L_{g_2} h v \geq 0\,.
\end{eqnarray}

\end{theorem}

\noindent{\bf Proof.} 
The pointwise minimization result is immediate from \eqref{eq-QPquad}--\eqref{eq-QP-DSSf}. 
For \eqref{C3eq2.4c}, \eqref{eq-1/2S} DSSf follows from 
\begin{align}
\dot h =\ & -\alpha(h(x))
+{1\over 2}\left( \omega +\sqrt{\omega^2
+\left(L_{g_2}h(L_{g_2}h)^{\T}\right)^2} \right)
\nonumber\\
& +|L_{g_1}h|\rho^{-1}(\max\{0, -h(x)\}) + L_{g_1}h d\,.
\end{align}\hfill $\Box$

\section{Inverse Optimal  Assignment of DSSf Gain}\label{C3sec3}

In the system \eqref{C3eq2.4c} there are two inputs: $u-u_0$ and $d$. This leads us to formulate the problem of safety filter design as a differential game \cite{basar-bernhard,doi:10.1137/1.9781611971132}. In this game, our control  $u-u_0$ wishes to remain small while keeping $h(x(t))$ from becoming too small. In contrast, the disturbance $d$ too wishes to remain small but its objective regarding $h(x(t))$ is to make it small and, in fact, negative. Since our goal in designing $u-u_0$ is to make the DSSf gain function from $d$ to the ``safety violation'' $-h(x)$ small, we refer to this problem as gain assignment. 

We pursue the following zero-sum two-player minimax (supinf, to be precise) optimization problem:
\begin{align}\label{C3eq8.5a}
&\sup_{u-u_0\in {\cal U}} \inf_{d\in {\cal D}}\bigg\{\lim_{t\rightarrow \infty}
\bigg[2\beta h(x(t))
+\int_0^t \bigg( l(x,u_0)
\nonumber\\ &
-(u-u_0)^{\T}R_2(x,u_0)(u-u_0)
+\beta\lambda\gamma\left(\frac{|d|}
{\lambda}\right)\bigg){\rm d}\tau \bigg]\bigg\},
\end{align}
where  ${\cal U}, {\cal D}$ are sets of locally bounded functions of $x$. In this problem, $R_2(x,u_0)=R_2(x, u_0)^{\T} > 0$ for all $x$ and $u_0$, $\gamma$ 
and $\gamma^{'}$ are in class ${\cal K}_\infty$, the  constants $\beta,\lambda>0$ are at the designer's disposal, and $l(x,u_0)$ is a weight on the state, upper bounded by a class ${\cal K}_\infty$ function of $h$. 
As indicated earlier, both $u-u_0$ and $d$ wish to remain small, however, $u-u_0$ wishes to keep $h$ positive (safe) while $d$ wishes to make $h$ negative (unsafe). 

We do not approach the game \eqref{C3eq8.5a} as a direct optimal control problem. Instead, we pursue it through {\em inverse optimality}. Both the control law $u-u_0$ and the weights $l(x,u_0), R_2(x,u_0), \gamma(\cdot)$ are up to the designer to choose. Even $h(x)$ is available for design, for a given safe set ${\cal C}$.

Before we continue, let us introduce the following notation:
For a class ${\cal K}_\infty$ function $\gamma$ whose derivative exists 
and is also a class ${\cal K}_\infty$ function, $\ell\gamma$ denotes the 
Legendre-Fenchel transform
\begin{align}
\ell\gamma (r)=& \int_0^r (\gamma^{'})^{-1}(s) d s\, \label{C3eq8.1b}
\\
=& r (\gamma^{'})^{-1}(r) -\gamma \left((\gamma^{'})^{-1}(r)
\right) \,, \ \ \mbox{(by Lemma~\ref{lema.3}.a)}\label{C3eq8.1a}
\end{align}
where $(\gamma^{'})^{-1}(r)$ stands for the inverse function of 
$\displaystyle{\frac{d \gamma(r)}{dr}}$. 

\begin{theorem}	\label{C3thm8.1} 
Consider the auxiliary system of (\ref{C3eq2.4})
\begin{equation} \label{C3eq8.2}
\dot{x} = f(x) - g_1(x)\ell \gamma(2|L_{g_1}h|)
\frac{(L_{g_1}h(x))^{\T}}{|L_{g_1}h|^2} + g_2(x) u 
\end{equation}
with a nominal control law $u_0(x,t)$, where 
$\gamma$ is a 
class ${\cal K}_\infty$ function whose derivative $\gamma^{'}$ is 
also a class ${\cal K}_\infty$ function. Suppose that, for a given $u_0$, there exists 
a matrix-valued function $R_2(x,u_0)=R_2(x,u_0)^{\T}>0$ such that the control law of the form
\begin{eqnarray}
u=u_0 + \bar u(x,u_0):= u_0+ R_2(x,u_0)^{-1} \left(L_{g_2} h\right)^{\T}  \label{C3eq8.3}
\end{eqnarray}
ensures safety of the system (\ref{C3eq8.2}) with respect to CBF candidate $h(x)$, namely, ensures that
\begin{align} \label{C3e8.7}
&L_{f+g_2 u_0}h  - \ell \gamma(2|L_{g_1}h|)
\nonumber\\ 
&+ L_{g_2}h R^{-1}_2 \left(L_{g_2}h\right)^{\T}
 \geq -\alpha(h(x)) 
\end{align}
for some $\alpha\in {\cal K}_h$. 
Then the control law
\begin{eqnarray}
u & = & u_0 + \bar u^*(x,u_0)
\ := \ u_0 +\beta\bar u(x,u_0)
\nonumber\\ 
&=&u_0+\beta R^{-1}_2 \left(L_{g_2}h\right)^{\T}\,, \quad \beta \geq 2
\label{C3eq8.4}
\end{eqnarray}
applied to (\ref{C3eq2.4}) 
{\em maximizes} the cost functional
\begin{eqnarray}\label{C3eq8.5}
J(u-u_0) &=& \inf_{d\in {\cal D}}\bigg\{\lim_{t\rightarrow \infty}
\bigg[2\beta h(x(t))
+\int_0^t \bigg( l(x,u_0)
\nonumber\\ &&
-(u-u_0)^{\T}R_2(x,u_0)(u-u_0)
\nonumber\\ &&
+\beta\lambda\gamma\left(\frac{|d|}
{\lambda}\right)\bigg){\rm d}\tau \bigg]\bigg\}
\end{eqnarray}
for any $\lambda \in (0, 2]$, where
\begin{eqnarray}
l(x, u_0) & = & -2 \beta \bigg[ L_{f+g_2u_0} h - \ell \gamma(2|L_{g_1}h|) 
\nonumber\\ &&
+ L_{g_2}h R^{-1}_2 \left(L_{g_2}h\right)^{\T}\bigg]\nonumber\\
& & -\beta(2-\lambda)\ell \gamma(2|L_{g_1}h|)
\nonumber\\ &&
-\beta(\beta-2) L_{g_2}h R^{-1}_2 \left(L_{g_2}h\right)^{\T}
\label{C3e8.6}
\\ 
&\leq& 2\beta \alpha(h(x))
\,\label{C3e8.6a}
\end{eqnarray}
is decrescent in the CBF $h$. 
\end{theorem}

\noindent {\bf Proof.} Thanks to (\ref{C3e8.7}), (\ref{C3e8.6}), we get (\ref{C3e8.6a}). 
Substituting $l(x)$ 
into (\ref{C3eq8.5}), it follows that 
\begin{eqnarray} \label{C3eq8.9}
&& J(u) 
\nonumber\\
& = & \inf_{d\in {\cal D}} \bigg\{\lim_{t\rightarrow \infty} 
\bigg[2\beta h(x(t)) 
\nonumber\\ && 
+\int_0^t \bigg(-2\beta L_{f+g_2 u_0}h  
+ \beta\lambda \ell \gamma(2|L_{g_1}h|)
\nonumber\\ && 
- \beta^2  L_{g_2}h R^{-1}_2\bigg(L_{g_2}h\bigg)^{\T} 
\nonumber\\ && 
-(u-u_0)^{\T}R_2(u-u_0)
+\beta\lambda \gamma\bigg(\frac{|d|}{\lambda}\bigg)\bigg) d\tau\bigg]
\bigg\} \nonumber\\
&= &\inf_{d\in {\cal D}} \bigg\{\lim_{t\rightarrow \infty} 
\bigg[2\beta h(x(t))
\nonumber\\ && 
-2\beta \int_0^t \bigg(L_{f+g_2 u_0}h  + L_{g_1}h d +L_{g_2}h (u-u_0)\bigg) 
d\tau  \nonumber\\
& & - \int_0^t \bigg( (u-u_0)^{\T}R_2(u-u_0)- 2\beta L_{g_2}h (u-u_0) 
\nonumber\\ && 
+ {\beta^2} L_{g_2}h R^{-1}_2 \bigg(L_{g_2}h\bigg)^{\T}\bigg)d\tau 
\nonumber\\
& & +\int_0^t \bigg(\beta\lambda\gamma\bigg(\frac{|d|}
{\lambda}\bigg)
+2\beta L_{g_1}h d 
\nonumber\\ && 
+\beta\lambda\ell \gamma(2|L_{g_1}h|)
\bigg)d\tau\bigg]\bigg\}\nonumber\\
& = &\inf_{d\in {\cal D}}\bigg\{\lim_{t\rightarrow \infty} 
\bigg[2\beta h(x(t))-2 \beta \int_0^t dh
\nonumber\\ && 
-\int_0^t (u-u_0-\bar u^*)^{\T}R_2 (u-u_0-\bar u^*) d\tau  \nonumber\\
& & +\beta\int_0^t \bigg[\lambda\gamma\bigg(\frac{|d|}{\lambda}\bigg)
-\lambda\gamma\bigg((\gamma^{'})^{-1}(2|L_{g_1}h|)\bigg) \nonumber\\
& &
+2\bigg(\lambda|L_{g_1}h| (\gamma^{'})^{-1}(2|L_{g_1}h|)
+L_{g_1}h d\bigg)\bigg] d\tau\bigg]\bigg\}
\nonumber\\ &&
\qquad\qquad\qquad\qquad\qquad\qquad {\mbox {(by (\ref{C3eq8.1a}))}} \nonumber\\
& = & 2 \beta h(x(0))
 +\beta\lambda \inf_{d\in {\cal D}}\int_0^{\infty}  \Pi(d,d^*) \, {\rm d} t
\nonumber\\
& & -\int_0^\infty (u-u_0-\bar u^*)^{\T}R_2 (u-u_0-\bar u^*) d\tau \, {\rm d} t
\,,
\end{eqnarray}
where 
\begin{eqnarray}
\Pi(d,d^*) &=& \gamma\bigg(\frac{|d|}{\lambda}\bigg)
-\gamma\bigg(\frac{|d^*|}{\lambda}\bigg) 
\nonumber\\ && 
- \gamma^{'}\bigg(\frac{|d^*|}{\lambda}\bigg)\frac{(d^*)^{\T}}
{\lambda|d^*|}(d^*-d)
\end{eqnarray}
and 
\begin{eqnarray}\label{C3eq8.10}
d^*(x)=-{\lambda}(\gamma^{'})^{-1}(2|L_{g_1}h|)
\frac{(L_{g_1}h)^{\T}}{|L_{g_1}h|}\,.
\end{eqnarray}
By Lemma~\ref{lema.3}.d, $\Pi(d,d^*)$ can be rewritten as 
\begin{eqnarray}\label{C3eq8.11}
\Pi(d,d^*) &=&  \gamma\left(\frac{|d|}{\lambda}\right)
+\ell\gamma\left(\gamma^{'}\left(\frac{|d^*|}{\lambda}\right)\right)
\nonumber\\ && 
+\gamma^{'}\left(\frac{|d^*|}{\lambda}\right)\frac{(d^*)^{\T}}{|d^*|}
\frac{d}{\lambda}\,.
\end{eqnarray}
Then by Lemma~\ref{lema.4} we have 
\begin{eqnarray}\label{C3eq8.12}
\Pi(d,d^*)&\geq& \gamma\left(\frac{|d|}{\lambda}\right)
+\ell\gamma\left(\gamma^{'}\left(\frac{|d^*|}{\lambda}\right)\right)
\nonumber\\ && 
-\gamma\left(\frac{|d|}{\lambda}\right)
-\ell\gamma\left(\gamma^{'}\left(\frac{|d^*|}{\lambda}\right)\right)
\nonumber\\ & 
=&0\,,
\end{eqnarray}
and $\Pi(d,d^*)=0$ if and only if 
$\displaystyle{\frac{d}{\lambda}
=(\gamma^{'})^{-1}\left(\gamma^{'}\left(\frac{|d^*|}{\lambda}\right)
\right)\frac{d^*}{|d^*|}}$, that is,
\begin{eqnarray}\label{C3eq8.12a}
\Pi(d,d^*)=0 \qquad \mbox{iff} \qquad d=d^*\,.
\end{eqnarray}
Thus 
\vspace{-2ex}
\begin{eqnarray}\label{C3eq8.12b}
\inf_{d\in {\cal D}} \int_0^\infty \Pi(d,d^*) \, {\rm d} t = 0\,,
\end{eqnarray}
and the ``worst case'' disturbance\index{``worst case'' disturbance} is given by (\ref{C3eq8.10}). The 
maximum of (\ref{C3eq8.9}) is reached with $u=u_0+\bar u^*$. Hence the 
control law (\ref{C3eq8.4}) maximizes the cost functional (\ref{C3eq8.5}). 
The value function of (\ref{C3eq8.5}) is $J^*(x)=2\beta h(x)$.
\hfill $\Box$

The parameter $\beta \geq 2$ in the statement of Theorem~\ref{C3thm8.1} 
represents a design degree of freedom. The parameter $\lambda$ 
(note that it parameterizes not only the penalty on the disturbance but 
also the penalty on the state's proximity to the boundary, i.e., the reward for the state's distance from the boundary, $l(x, u_0)$) indicates that the same control 
law is inverse optimal with respect to an entire family of different 
cost functionals. 

\begin{remark}\label{C3rem8.1}
\em One approach to studying safety in the presence of inputs and disturbances is  reachability  \cite{8263977,choi2021robust,YIN2020104736,9126836}, where {\em Hamilton-Jacobi-Isaacs} (HJI) PDEs
arise and need to be solved. Even though not explicit in the proof of Theorem~\ref{C3thm8.1}, 
the CBF $h(x)$ solves the following family of HJI equations:
\begin{eqnarray}
 L_{f+g_1u_0}h +\frac{\beta}{2} L_{g_2}h R_2(x,u_0)^{-1}\left(L_{g_2}h\right)^{\T}
-\frac{\lambda}{2}\ell \gamma(2|L_{g_1}h|)
\nonumber\\ 
+\frac{l(x,u_0)}{2\beta}= 0 \,, \label{C3eq8.13}
\end{eqnarray}
parameterized by $(\beta, \lambda)\in [2, \infty)\times(0,2]$. 
\hfill $\Box$
\end{remark}

\begin{remark}\label{rem-atten}\em
It is also easily seen from the proof of Theorem~\ref{C3thm8.1} that, even for  initial 
conditions  on the boundary, the achieved disturbance attenuation\index{disturbance attenuation} level is 
\begin{align}\label{C3eq8.14}
&2\beta h(x(t))+ 2\beta\int_0^\infty  \alpha(h(x)) \, {\rm d} t \geq 2\beta h(x(t))+\int_0^\infty  l(x,u_0) \, {\rm d} t 
\nonumber\\ 
&\geq  
\int_0^\infty (u-u_0)^{\T}R_2(x,u_0) (u-u_0) \, {\rm d} t  
-\beta\lambda\int_0^\infty \gamma\left(\frac{|d|}{\lambda}\right) \, {\rm d} t
\nonumber\\ 
&\geq  
-\beta\lambda\int_0^\infty \gamma\left(\frac{|d|}{\lambda}\right) \, {\rm d} t\,.
\end{align}
Summarizing, we refer to the property 
\begin{align}\label{C3eq8.14a}
 h(x(t))+ \int_0^\infty  \alpha(h(x)) \, {\rm d} t \geq
-{\lambda\over 2}\int_0^\infty \gamma\left(\frac{|d|}{\lambda}\right) \, {\rm d} t\,,
\end{align}
as {\em integral disturbance-to-state safety} (iBSSf). 
\hfill $\Box$
\end{remark}

\begin{example}\label{C3exam8.1a}\em
Consider the system from Example \ref{ex-1+x2d}. Take $\gamma(r) = \ell \gamma(2r) = r^2$. With 
\begin{equation}
R_2 = {1\over \max\left\{ 0, u_0 -\alpha(h(x))\right\}+(1+x^2)^2}>0\,,
\end{equation} 
condition \eqref{C3e8.7} is satisfied. The control \eqref{C3eq8.4} is given by 
\begin{equation}\label{eq-ex3cont}
u = u_0 - {\beta\over R_2} = u_0 + \beta\left[\min\left\{ 0, -u_0 +\alpha(h(x))\right\}-(1+x^2)^2\right]
\end{equation}
and, for all $\beta \geq 2$, is the maximizer of 
\begin{eqnarray}\label{C3eq8.5dex}
J(u-u_0) &=& \inf_{d\in {\cal D}}\bigg\{\lim_{t\rightarrow \infty}
\bigg[2\beta x(t)
+\int_0^t \bigg( l(x,u_0)
\nonumber\\ &&
-R_2{(u-u_0)^2}
+{\beta\over \lambda} d^2\bigg){\rm d}\tau \bigg]\bigg\}
\end{eqnarray}
for any $\lambda \in (0, 2]$, with $l(x,u_0) \leq -2\beta x$, and achieves
\begin{align}\label{C3eq8.14ex}
&x(+\infty)+ \int_0^\infty  x(t) \, {\rm d} t 
\leq{1\over 2\lambda} \int_0^\infty d^2(t) \, {\rm d} t\,. 
\end{align}
And, $\forall\beta\geq 1$,  controller \eqref{eq-ex3cont} with $\alpha(h)=h$ guarantees
\begin{equation}
x(t) \leq {\rm e}^{-t}x_0 +{1\over 4}\left(\sup_{0\leq \tau\leq t} |d(\tau)|\right)^2\,, \quad \forall t\geq 0\,.
\end{equation}
\hfill $\Box$
\end{example}

\section{Is DSSf QP Safety Filter Inverse Optimal?}\label{sec-DSSf-invopt}

The following theorem, proven by mimicking the proof of Theorem \ref{C3thm8.1}, answers this section's question in the affirmative. 

\begin{theorem}	\label{C3thm8.1d} 
Consider  system \eqref{C3eq2.4c} with  associated DSSf-CBF $h$ and a gain function $\rho$. 
For any $\beta \geq 2$, the control law 
\begin{eqnarray}
u & = & u_0 + \bar u_{\rm QP}^*(x,u_0)=u_0 +\beta \bar u_{\rm QP}(x,u_0)\,,\label{C3eq8.4d}
\end{eqnarray}
with $\bar u_{\rm QP}$ defined in \eqref{eq-QP-DSSf} and $\omega$ defined in \eqref{C3eq8.19nom},  
{\em maximizes} 
\begin{align}\label{C3eq8.5d}
J(u-u_0) =& \inf_{d\in {\cal D}}\bigg\{\lim_{t\rightarrow \infty}
\bigg[2\beta h(x(t))
+\int_0^t \bigg( l(x,u_0)
\nonumber\\ &
-R_2(x,u_0)|u-u_0|^2 +{\beta\over\lambda} R_1(x) |d|^2
\bigg){\rm d}\tau \bigg]\bigg\} 
\end{align}
for all $ \lambda \in (0, 2]$, where
\begin{align}
\label{eq-R1-QP}
R_1(x) =\ & {1\over \rho^{-1}(\max\{0, -h(x)\})} >0
\\ \label{eq-R2-QP}
R_2(x,u_0) =\  & {|L_{g_2}h|^2 \over \max\left\{0,-\omega\right\}} >0 
\\ \label{eq-l-QP}
l(x,u_0) =\ & 2 \beta\alpha(h(x)) + \beta\min\{\beta \omega, - 2\omega\}
\nonumber\\
\ &-\beta(2-\lambda) \left| L_{g_1}h\right| \rho^{-1}(\max\{0, -h(x)\})
\nonumber \\
\leq\ & 2 \beta\alpha(h(x))\,.
\end{align}
\end{theorem}

The weight $R_1$ in \eqref{eq-R1-QP} is infinite in the safe set $h(x) \geq 0$ since the ``optimal disturbance''
\begin{equation}
d^*(x) = - \lambda \rho^{-1}(\max\{0, -h(x)\})\frac{(L_{g_1}h(x))^{\T}}{|L_{g_1}h|}
\end{equation}
wastes no effort in the safe set. Likewise, $R_2$ in \eqref{eq-R2-QP} is infinite when $\omega\geq 0$ since  control \eqref{eq-QP-DSSf} puts in no  effort when $u_0$ makes the system safe on its own. We also recall from Remark \ref{rem-Lg2h=0} that \eqref{eq-QP-DSSf} precludes $L_{g_2} h$ from being zero when $\omega <0$, so $R_2$ can, in fact, never be zero, namely, $u-u_0$ is penalized for all $x$. 

\begin{example}\label{ex-1+x2d+}\em
Back to Ex. \ref{ex-1+x2d},  control $u= u_0 + \beta \bar u_{\rm QP}$, $\beta\geq 2$, with $\bar u_{\rm QP} = \min\left\{ 0, -u_0 -\rho^{-1}(\max\{0, x\})+x\right\}$, results in
\begin{align}\label{C3eq8.14ex+}
&x(+\infty)+ \int_0^\infty  x(t) \, {\rm d} t 
\leq{1\over 2\lambda} \int_0^\infty {d^2(t)\over \rho^{-1}(\max\{0, x(t)\})} \, {\rm d} t\,, 
\end{align}
which, unlike control \eqref{eq-ex3cont} in Example \ref{C3exam8.1a}, fails to achieve a finite integral gain in the safe set $x\leq 0$ like \eqref{C3eq8.14ex}. 
\hfill $\Box$
\end{example}

\section{Inverse Optimal QP Safety Filter for Disturbance-Free Systems}\label{sec-disturbance-free}

In the disturbance-free system, 
\begin{equation} \label{C3eq2.4e}
\dot x = f (x) +g_2 (x) u_0 + g_2 (x) (u-u_0)\,,
\end{equation}
let us introduce
\begin{eqnarray}\label{C3eq8.19free}
\omega_2(x,u_0) = L_{f+g_2 u_0}h   +\alpha(h(x))\,
\end{eqnarray}
and the control law
$u = u_0 + \bar u_{\rm QP2}(x,u_0)$
with (recalling Remark \ref{rem-Lg2h=} on 
division by $L_{g_2}h = 0$)
\begin{equation}\label{eq-QP-DSSf2}
\bar u_{\rm QP2} 
= (L_{g_2}h)^T \displaystyle{\max\left\{0, -\omega_2(x,u_0)\right\}\over |L_{g_2}h|^2}\,.
\end{equation}
This standard QP safety filter  renders  system \eqref{C3eq2.4e} safe w.r.t.  CBF $h(x)$. 
The next result follows from Theorem \ref{C3thm8.1d}.

\begin{corollary}	\label{C3thm8.1e} 
For  system \eqref{C3eq2.4e} and  any $\beta \geq 2$, the control  
\begin{eqnarray}
u & = & u_0 + \bar u_{\rm QP2}^*(x,u_0)=u_0 +\beta \bar u_{\rm QP2}(x,u_0)\,,\label{C3eq8.4e}
\end{eqnarray}
with $\bar u_{\rm QP2}$ defined in \eqref{eq-QP-DSSf2}, {\em maximizes} the cost functional
\begin{eqnarray}\label{C3eq8.5e}
J(u-u_0) &=& \lim_{t\rightarrow \infty}
\bigg[2\beta h(x(t))
+\int_0^t \bigg( l(x,u_0)
\nonumber\\ &&
- 
{|L_{g_2}h|^2 |u-u_0|^2\over\max\left\{0,-\omega_2\right\} }
\bigg){\rm d}\tau \bigg]\
\end{eqnarray}
 with $l(x,u_0) \leq 2\beta \alpha(h)$. 
\end{corollary}

\section{Inverse Optimality with Reciprocal CBFs}\label{sec-RCBF}\label{sec-disturbance-freeR}

It is worth examining inverse optimality in terms of Reciprocal CBFs (RCBFs) of the form $B(x) = 1/h(x)$, without designing any new safety filters but merely recast inverse optimality in the context of RCBFs. 


\begin{theorem}	\label{C3thm8.1B} 
Suppose that, for a given $u_0$, there exists 
a matrix-valued function $\bar R(x,u_0)=\bar R(x,u_0)^{\T}>0$ such that the control law of the form
\begin{eqnarray}
u=u_0 + \bar u(x,u_0):= u_0- \bar R(x,u_0)^{-1} \left(L_{g_2} B\right)^{\T}  \label{C3eq8.3B}
\end{eqnarray}
ensures safety of the system (\ref{C3eq2.4e}) with respect to an RCBF candidate $B(x)$, namely, ensures that
\begin{align} \label{C3e8.7B}
&L_{f+g u_0} h - L_{g}B \bar R^{-1} \left(L_{g}B\right)^{\T}
 \leq \bar\alpha\left({1\over B(x)}\right) 
\end{align}
for some $\bar\alpha$ in class $ {\cal K}$. 
Then the control law
\begin{eqnarray}
u & = & u_0 + \bar u^*(x,u_0)
\ := \ u_0 +\beta\bar u(x,u_0)
\nonumber\\ 
&=&u_0-\beta \bar R^{-1} \left(L_{g}B\right)^{\T}\,, \quad \beta \geq 2\,,
\label{C3eq8.4B}
\end{eqnarray}
applied to system (\ref{C3eq2.4e}), {\em minimizes} the cost functional
\begin{eqnarray}\label{C3eq8.5B}
J(u-u_0) &=& \lim_{t\rightarrow \infty}
\bigg[-{2\beta\over B(x(t))}
+\int_0^t \bigg(\bar l(x,u_0)
\nonumber\\ &&
+(u-u_0)^{\T}\bar R(x,u_0)(u-u_0)
\bigg){\rm d}\tau \bigg]\
\end{eqnarray}
for any $\lambda \in (0, 2]$, where
\begin{eqnarray}
\bar l(x, u_0) & = & -2 \beta \bigg[ L_{f+g_2u_0} h 
+ L_{g}B \bar R^{-1} \left(L_{g}B\right)^{\T}\bigg]\nonumber\\
& & 
+\beta(\beta-2) L_{g}B \bar R^{-1} \left(L_{g}B\right)^{\T}
\label{C3e8.6R}
\\ 
&\geq& -{2\beta\over B^2(x(t))} \bar\alpha\left({1\over B(x)}\right)
\,\label{C3e8.6aR}
\end{eqnarray}
is lower bounded by a negative definite function of 
$1/B(x)$. 
\end{theorem}

\noindent{\bf Proof.} 
By adapting the proof of Theorem \ref{C3thm8.1} to the case $g_1=0, g_2=g$, with $ l= -\bar l$, $R_2 = h^2 \bar R $, and $\alpha(h) = h^2 \bar\alpha(h)$. 
\hfill $\Box$

It is of interest to return to the QP filter $u = u_0 + \bar u_{\rm QP2}$ with \eqref{eq-QP-DSSf2} and \eqref{C3eq8.19free}, which  can be expressed as 
\begin{equation}\label{eq-QPB}
u = u_0 + \tilde u_{\rm QP}(x,u_0)
\end{equation}
with (recalling  Remark \ref{rem-Lg2h=}  on division by $L_g B =0$)
\begin{eqnarray}\label{eq-QP-DSSfsfB}
\tilde u_{\rm QP} &=& -(L_{g}B)^T
\displaystyle{\max\left\{0, \tilde\omega(x,u_0)\right\}\over |L_{g}B|^2} 
\\ 
\label{C3eq8.19freeB}
\tilde\omega(x,u_0) &=& L_{f+g u_0} B  -\bar \alpha\left({1\over B(x)}\right)\,.
\end{eqnarray}
With Theorem \ref{C3thm8.1B}, we restate Corollary \ref{C3thm8.1e}.

\begin{corollary}\label{C3thm8.1eB}
Let a function $B$ be such that $h=1/B$ satisfies Definition \ref{def-CBF},  and let $B$ be a reciprocal CBF, namely, let $B$ satisfy the condition 
\begin{equation}
L_g B = 0 \quad \Rightarrow \quad \tilde\omega \leq 0\,,
\end{equation}
 with $\tilde\omega$ defined in \eqref{C3eq8.19freeB}. Then the safety filter
\begin{equation}\label{eq-QPBbeta}
u = u_0 + \beta \tilde u_{\rm QP}(x,u_0)\,, \quad\beta\geq 2
\end{equation}
with $\tilde u_{\rm QP}$ defined in \eqref{eq-QP-DSSfsfB}, {\em minimizes}  \eqref{C3eq8.5B} with
\begin{equation}\label{eq-R2-QP3B}
\bar R(x,u_0) = {|L_{g}B|^2 \over\max\left\{0,\tilde\omega\right\} }>0\,
\end{equation}
and $\bar l(x, u_0) \geq -{2\beta\over B^2(x(t))} \bar\alpha\left({1\over B(x)}\right)$. 
\end{corollary}

\section{Stochastic CBFs}\label{sec-stochCBF}

We return to the barrier function in Definition \ref{def-CBF}, under Assumption \ref{ass-setC}, but now consider the stochastic system 
\begin{eqnarray}
{\rm d} x=f (x )\, {\rm d} t+g_1(x )\, {\rm d} w,\label{eqn:gs}
\end{eqnarray}
where 
$w$ is an $r$-dimensional independent standard 
Wiener process\index{Wiener process}
and $f,g$ 
are locally Lipschitz.

For the barrier function candidate $h(x)$, we recall that It\^{o}'s lemma states that
\begin{equation}
{\rm d}h = {\cal L} h \, {\rm d} t + L_{g_1}h \, {\rm d} w
\end{equation}
where
\begin{equation}
{\cal L} h = L_f h +\frac{1}{2} {\rm Tr}\left\{g_1^{\T}{{\partial^2h}\over{\partial x^2}}g_1\right\}  \,
\end{equation}
is referred to as the {\em infinitesimal generator} of $h$. 

We say that the system \eqref{eqn:gs} satisfies the {\em stochastic barrier function condition} (SBFc) if
there exists a function $\alpha\in{\cal K}_h$ 
such that, for all $x\in \IR^n$, the following function is nonnegative,
\begin{equation}\label{eq-stochomega}
\omega(x) = L_f V +\frac{1}{2} {\rm Tr}\left\{g_1^{\T}{{\partial^2h}\over{\partial x^2}}g_1\right\} +\alpha(h)\,.
\end{equation}

From here on we proceed formally, with systems and controllers that satisfy the SBFc, without going a step further to establish safety in probability, or at least in the mean, which would be done by employing the techniques as in the proof of Theorem 3.2 in \cite{Krstic-Deng-book}, the techniques in Theorem 3 in \cite{CLARK2021Stochastic}, or the technique in the proof of Lemma 1 in \cite{WuquanStochasticNonovershooting}.

Now we turn our attention to systems that, in addition to the noise input $w$, have a control input $u\in \IR^{m_2}$:
\begin{eqnarray}
\, {\rm d} x=f (x )\, {\rm d} t+g_1 (x )\, {\rm d} w+g_2 (x )u\, {\rm d} t\,.\label{eqn:su}
\end{eqnarray}

\begin{definition}	\label{C3def2.1s}
A scalar differentiable function $h$ is called a {\em stochastic control barrier 
function (SCBF)}\ for (\ref{eqn:su}) if there exists a function $\alpha\in{\cal K}_{(
0,\sup h(\xi))}$ such that the following implication 
holds for all $x\in\left\{\eta \in\IR^n \left| 
0\leq h(\eta) < \sup_{\xi \in\IR^n} h(\xi)\right. \right\}$:
\begin{equation} \label{C3eq2.5s}
\displaystyle{\sup_{u\in\IR^{m_2}}\left\{L_fV+{1\over 2}{\rm Tr}\left\{g_1^{\T}{{\partial^2h}\over{\partial x^2}}
g_1\right\}+L_{g_2}hu \right\}}  \geq  - \alpha(h)
\end{equation}
\end{definition}

The following  is obtained by adapting \cite[Lemma 2.1]{661589}.

\begin{lemma}\label{C3lem2.1s}
A function $h$ is an SCBF, namely, it satisfies (\ref{C3eq2.5s}) in Definition~\ref{C3def2.1s} if and only if
\begin{equation}
L_{g_2}h =0 \quad \Rightarrow \quad \omega \geq 0 \,,
\end{equation}
where $\omega(x)$ is defined in \eqref{eq-stochomega}.
\end{lemma}

Next, a Sontag-type control law ensures stochastic safety. 

\begin{theorem}\label{thm-basic-stoch-QP}
Under the control law \eqref{C3eq8.18}, \eqref{C3eq8.18kappa} with $\omega(x)$ defined in \eqref{eq-stochomega}, the system \eqref{eqn:su} satisfies the SBFc. 
\end{theorem}

\noindent{\bf Proof.} 
A direct substitution yields
\begin{eqnarray}
{\cal L} h =L_f h +\frac{1}{2} {\rm Tr}\left\{g_1^{\T}{{\partial^2h}\over{\partial x^2}}g_1\right\} + L_{g_2} h u_{\rm S} \geq -\alpha(h(x)) \,.
\end{eqnarray}
\hfill $\Box$

\section{Inverse Optimal Stochastic Safety Filters}\label{sec-stochinvop}

Next, we turn our attention to systems where, in addition to ensuring safety in the presence of noise $w$, the task of control input $u$ is to stick close to the nominal $u_0$: 
\begin{equation}
\, {\rm d} x=[f (x )+g_2(x)u_0]\, {\rm d} t+g_1 (x )\, {\rm d} w+g_2 (x )(u-u_0)\, {\rm d} t\,.\label{eqn:sus}
\end{equation}

\begin{theorem}\label{thm:inop}
Consider the control law
\begin{eqnarray}
u&=& u_0 + \bar u(x,u_0)\\
\bar u(x,u_0) &=&R_2^{-1}\left(L_{g_2}h\right)^{\T}
{{\ell\gamma_2\left(\left |L_{g_2}hR_2^{-{1/2}}\right |\right)}\over{\left |L_{g_2}hR_2^{-{1/2}}\right |^2}},\label{eqn:clu}
\end{eqnarray}  
where $\gamma_2$ is a class ${\cal K}_\infty$
function whose derivative is also a class ${\cal K}_\infty$ function, and $R_2 (x )$ is a 
matrix-valued function such that $R_2(x)=R_2(x)^{\T}>0$. If the control law (\ref{eqn:clu}) 
makes the system (\ref{eqn:sus}) satisfy the SBFc with  
respect to an SCBF candidate $h (x )$, namely, if the following condition holds, 
\begin{align}\label{eq-SBFc*}
&L_{f+g_2 u_0}h +{1\over 2}{\rm Tr}\left\{g_1^{\T}{{\partial^2h}\over{\partial x^2}}g_1\right\}
+\ell\gamma_2\left(\left |L_{g_2}hR_2^{-{1/2}}\right |\right) 
\nonumber\\
& \geq - \alpha(h)\,,
\end{align}
then the control law
\begin{align}
u=\ & u_0 + \bar u^*(x,u_0)\\
\bar u^*=\ &{\beta\over 2}R_2^{-1}\left(L_{g_2}h\right)^{\T}
{{\left(\gamma_2'\right)^{-1}\left(\left |L_{g_2}hR_2^{-{1/2}}\right |\right)}\over{\left |L_{g_2}hR_2^{-{1/2}}\right |}},\;\;\beta\geq 2\label{eqn:clu*}
\end{align}  
also makes the system (\ref{eqn:sus}) satisfy the SBFc and, moreover, {\em maximizes} the cost functional
\begin{align}
J (u -u_0)=&\lim_{t\to\infty}E\bigg\{2\beta h (x (t ) ) 
+ \int_0^t \bigg[l (x , u_0)\nonumber\\
&-\beta^2\gamma_2\left({2\over\beta}\left |R_2^{1/2}(u-u_0)\right |\right)\bigg]d\tau\bigg\},\label{eqn:opf}
\end{align}
where
\begin{eqnarray}
l (x, u_0 )&=&-2\beta \bigg[
L_{f+{g_2}u_0}h +{1\over 2}{\rm Tr}\left\{g_1^{\T}{{\partial^2h}\over{\partial x^2}}g_1\right\}
\nonumber\\
&& +\ell\gamma_2\left(\left |L_{g_2}hR_2^{-{1/2}}\right |\right)
\bigg]\nonumber\\
&&-\beta \left(\beta -2\right)\ell\gamma_2\left(\left |L_{g_2}hR_2^{-{1/2}}\right |\right)
\nonumber\\
&\leq & 2\beta\alpha(h)\,.
\end{eqnarray}
\end{theorem}

\noindent{\bf Proof.}
Before we engage into proving that the control law (\ref{eqn:clu*}) minimizes (\ref{eqn:opf}), we first show that makes the system (\ref{eqn:sus}) satisfy the SBFc. With Lemma \ref{lema.3} we get
\begin{eqnarray}
{\cal L}h\mid_{(\ref{eqn:clu*})}
&=&L_{f+g_2 u_0}h+{1\over 2}{\rm Tr}\left\{g_1^{\T}{{\partial^2h}\over{\partial x^2}}g_1\right\}
\nonumber\\
&& +{\beta\over 2}\left |L_{g_2}hR_2^{-{1/2}}\right |\left(\gamma_2'\right)^{-1}\left(\left |L_{g_2}hR_2^{-{1/2}}\right |\right)\nonumber\\
&=&L_{f+g_2 u_0}h+{1\over 2}{\rm Tr}\left\{g_1^{\T}{{\partial^2h}\over{\partial x^2}}g_1\right\}
\nonumber\\
&&+{\beta\over 2}\left[\ell\gamma_2\left(\left |L_{g_2}hR_2^{-{1/2}}\right |\right)\right.\nonumber\\
&&\left.+\gamma_2\left(\left(\gamma_2'\right)^{-1}
\left(\left |L_{g_2}hR_2^{-{1/2}}\right |\right)\right)\right]\nonumber\\
&\geq&{\cal L}h\mid_{(\ref{eqn:clu})} \geq - \alpha(h) \,,
\end{eqnarray}
which proves that (\ref{eqn:clu*}) makes the system (\ref{eqn:sus}) satisfy the SBFc.

Now we prove optimality. Recalling that the It\^{o} differential of $h$ is 
\begin{eqnarray}
{\rm d}h={\cal L}h (x )\, {\rm d} t+{{\partial h}\over{\partial x}}g_1 (x )\, {\rm d} w,
\end{eqnarray}
according to the property of It\^{o}'s integral\index{It\^{o}'s integral}~\cite[Theorem 3.9]{¿ksendal2010stochastic}, we get
\begin{eqnarray}
E\left\{h (0 )-h (t )+\int_0^t{\cal L}h (x (\tau ) )d\tau\right\}=0.
\end{eqnarray}
Then substituting $l (x )$ into $J (u )$, we have
\begin{align}
&J (u -u_0)
\nonumber\\
=&\lim_{t\to\infty}E\bigg\{2\beta h (x (t ) ) 
+ \int_0^t \bigg[l (x,u_0 )\nonumber\\
&-\beta^2\gamma_2\left({2\over\beta}\left |R_2^{1/2}(u-u_0)\right |\right)\bigg]d\tau\bigg\}\nonumber\\
=&2\beta E\left\{h (x (0 ) )\right\}
+\lim_{t\to\infty}E\left\{\int_0^t\left[2\beta {\cal L}h\mid_{(\ref{eqn:sus})}
\right.\right. \nonumber \\ 
& \left.\left.+l (x,u_0 )
-\beta^2\gamma_2\left({2\over\beta}\left |R_2^{1/2}(u-u_0)\right |\right)\right]d\tau\right\}\nonumber\\
=&2\beta E\left\{h (x (0 ) )\right\}
\nonumber\\
&+\lim_{t\to\infty}E\left\{\int_0^t\left[-\beta^2\gamma_2\left({2 \over \beta}
\left |R_2^{1/2}(u-u_0)\right |\right)
\right.\right.\nonumber\\
&\left.\left.-\beta^2\ell\gamma_2\left(\left |L_{g_2}hR_2^{-{1/2}}\right |\right)
+2\beta L_{g_2}h(u-u_0)\right]d\tau\right\}.
\end{align}
Now we note that
\begin{eqnarray}
\gamma_2'\left({2\over\beta}\left |R_2^{1/2}\bar u^*\right |\right)
&=&\left |L_{g_2}hR_2^{-{1/2}}\right |\label{eqn:l2},
\end{eqnarray}
which yields
\begin{align}
&J (u-u_0 )
\nonumber\\
=&\lim_{t\to\infty}E\left\{\int_0^t
\left[-\beta^2\gamma_2\left (\left |{2\over\beta}R_2^{1/2}(u-u_0)\right |\right )
\right.\right.\nonumber\\
&\left.\left.-\beta^2\ell\gamma_2\left(\gamma_2'\left (\left |{2\over\beta}R_2^{1/2}\bar u^*\right |\right )\right)\bar u^*\right.\right.\nonumber\\
&\left.\left.+2\beta \gamma_2'\left (\left |{2\over\beta}R_2^{1/2}\bar u^*\right |\right ){{\left({2\over\beta}R_2^{1/2}\bar u^*\right)^{\T}}\over
{\left |{2\over\beta}R_2^{1/2}\bar u^*\right |}}R_2^{1/2}(u-u_0)\right]d\tau\right\}\nonumber\\
&+2\beta E\left\{h (x (0 ) )\right\}.
\end{align}
With the general Young inequality\index{Young's inequality} (Lemma \ref{lema.4}), we obtain
\begin{align}
&J (u-u_0 )\nonumber\\
\leq &2\beta E\left\{h (x (0 ) )\right\}+\lim_{t\to\infty}E\left\{\int_0^t
\left[-\beta^2\gamma_2\left (\left |{2\over\beta}R_2^{1/2}(u-u_0)\right |\right )
\right.\right.\nonumber\\
&-\beta^2\ell\gamma_2\left(\gamma_2'\left (\left |{2\over\beta}R_2^{1/2}\bar u^*\right |\right )\right)
\nonumber\\
&+\beta^2\gamma_2\left (\left |{2\over\beta}R_2^{1/2}(u-u_0)\right |\right )
\nonumber\\
&\left.\left.+\beta^2\ell\gamma_2\left(\gamma_2'\left (\left |{2\over\beta}
R_2^{1/2}\bar u^*\right |\right )\right)\right]d\tau\right\}
\nonumber\\
=&2\beta E\left\{h (x (0 ) )\right\},
\end{align}
where the equality holds if and only if
\begin{align}
&\gamma_2'\left (\left |{2\over\beta}R_2^{1/2}\bar u^*\right |\right )
{{\left({2\over\beta}R_2^{1/2}\bar u^*\right)^{\T}}\over{\left |{2\over\beta}R_2^{1/2}\bar u^*\right |}}
\nonumber\\
&=\gamma_2'\left (\left |{2\over\beta}R_2^{1/2}(u-u_0)\right |\right )
{{\left({2\over\beta}R_2^{1/2}(u-u_0)\right)^{\T}}\over{\left |{2\over\beta}R_2^{1/2}(u-u_0)\right |}},
\end{align}
that is, when $u-u_0=\bar u^*$. Thus
\begin{eqnarray}
\mbox{arg}\max_{u-u_0}J (u-u_0 )&=&\bar u^*\\
\max_{u-u_0} J (u-u_0 )&=&2\beta E\left\{h (x (0 ) )\right\}.
\end{eqnarray}
\mbox{} \hfill $\Box$

\begin{remark}\label{C3rem8.1s}{\rm 
Similar to Remark \ref{C3rem8.1}, even though not explicit in the proof of Theorem \ref{thm:inop}, $h (x )$ solves the following
family of {\it Hamilton-Jacobi-Bellman} equations
\index{Hamilton-Jacobi-Bellman equations} parameterized by $\beta\in [2,\infty)$:
\begin{eqnarray}
L_{f+g_2u_0}h+{1\over 2}{\rm Tr}\left\{g_1^{\T}{{\partial^2h}\over{\partial x^2}}g_1\right\}
+{\beta \over 2}\ell\gamma_2\left(\left |L_{g_2}hR_2^{-{1/2}}\right |\right)
\nonumber\\
+{l (x,u_0 )\over{2\beta}}=0.
\end{eqnarray}
}\end{remark}\mbox{} \hfill $\Box$

Theorem \ref{thm:inop} establishes inverse optimality but does not  design a controller that meets  condition \eqref{eq-SBFc*}. We pursue  an inverse optimal safety-ensuring control next, using QP. 

\begin{theorem}	\label{C3thm8.1es} 
For the system \eqref{eqn:sus} and for any $\beta \geq 2$, the control law 
\begin{eqnarray}
u & = & u_0 + \bar u_{\rm QP2}^*(x,u_0)=u_0 +\beta \bar u_{\rm QP2}(x,u_0)\label{C3eq8.4estoch}
\end{eqnarray}
employing a standard QP safety filter
\begin{equation}\label{eq-QP-DSSf2s}
\bar u_{\rm QP2} = (L_{g_2}h)^T
\displaystyle{\max\left\{0, -\omega(x,u_0)\right\}\over |L_{g_2}h|^2} 
\end{equation}
with
\begin{eqnarray}\label{C3eq8.19s}
\omega_2(x,u_0) = L_{f+g_2 u_0}h   +\frac{1}{2} {\rm Tr}\left\{g_1^{\T}{{\partial^2h}\over{\partial x^2}}g_1\right\}+\alpha(h(x))\,
\end{eqnarray}
{\em maximizes} the cost functional
\begin{align}\label{C3eq8.5es}
J(u-u_0) =& \lim_{t\rightarrow \infty}
E\bigg\{2\beta h(x(t))
+\int_0^t \bigg( l(x,u_0)
\nonumber\\ &
-{|L_{g_2}h|^2 |u-u_0|^2  \over\max\left\{0,-\omega_2\right\} }
\bigg){\rm d}\tau \bigg\}\
\end{align}
for any $\lambda \in (0, 2]$ and with some $l(x,u_0) \leq 2\beta \alpha(h)$. 
\end{theorem}

\noindent{\bf Proof.}
By verifying that \eqref{eq-SBFc*} is met with $\gamma_2(r) = {1\over 4}r^2$. 
\mbox{} \hfill $\Box$

\begin{example}\label{C3exam8.1as}\em
While a stochastic disturbance doesn't always have a detrimental effect on the SBFc condition, we construct an example in which the stochastic effect is indeed detrimental and where a QP safety filter acts  to mitigate this effect. Consider the system 
\begin{eqnarray}
{\rm d} x & = & u\,{\rm d}t+(1-x)\ {\rm d}w \,.\label{C3exa8.1as}
\end{eqnarray}
Let us take $u_0=0$, $h(x) = \ln(1-x)$, and $\alpha(h)=h$. We obtain that \eqref{C3eq8.19s} gives
$\omega_2(x,u_0) = -{1\over 2}+\ln(1-x)$, 
which yields \eqref{eq-QP-DSSf2s} in the form
\begin{equation}\label{eq-QP-DSSf2sex}
\bar u_{\rm QP2} = (x-1) \max\left\{0, {1\over 2}-\ln(1-x)\right\}
\end{equation}
The safety filter kicks in when $x>1-\sqrt{\rm e}$, which is negative. If the stochastic disturbance $w$ were absent, the control would be $\bar u_{\rm QP2} = (x-1) \max\left\{0, -\ln(1-x)\right\}$ and the safety filter would kick in only at $x=0$, i.e., never if $x_0<0$. 
\end{example}

\section{Noise-to-State Safety Filters}\label{sec-stochinvopNS}

In Sections \ref{sec-stochCBF} and \ref{sec-stochinvop} we studied stochastic systems of the form \eqref{eqn:su} with a known, unity covariance. This is quite limiting, regardless of the unity-intensity noise being the standard in stochastic optimal control. A stochastic disturbance acting on a system may be of unknown and time-varying incremental covariance $\Sigma(t)\Sigma(t)^{\T}\, {\rm d} t$, 
i.e., 
\begin{equation}\label{eq-Sigma}
E\left\{\, {\rm d} w\, {\rm d} w^{\T}\right\}=\Sigma(t)\Sigma(t)^{\T}\, {\rm d} t
\end{equation}
where $\Sigma(t)$ is a bounded function taking values in the set of nonnegative
definite matrices. For matrices $X=[x_1,x_2,\cdots,x_n]$, we use the Frobenius norm
\begin{eqnarray}
|X|_{\cal F}\eqdef\left({\rm Tr}\left\{X^{\T}X\right\}\right)^{1/2}
=\left({\rm Tr}\left\{XX^{\T}\right\}\right)^{1/2}\,
\end{eqnarray}
and note that $|X|_{\cal F}=|{\rm col}(X)|$, where ${\rm col}(X)=[x_1^{\T},x_2^{\T},\cdots,x_n^{\T}]^{\T}$.

When the covariance is unknown and time-varying, it needs to be treated as a deterministic disturbance in Sections \ref{sec-DSSf}, \ref{sec-DSSf-CBF}, \ref{sec-DSSf-filter}, and \ref{sec-DSSf-invopt}. Accordingly, only a graceful degradation of safety in the presence of the disturbance $\Sigma(t)$ can be expected, as in \eqref{C3eq2.5b}. We refer to such a stochastic property as {\em noise-to-state safety} (NSSf). However, we don't conduct analysis of achieving such a property in probability or in the mean. We just pursue the attainment of the following condition
\begin{equation}\label{eq-NSBFc}
\begin{array}{rcl}
&{\displaystyle  \min\left\{ 0, h(x)\right\}  \leq-\rho\left(\left|\Sigma\Sigma^{\T}\right|_{\cal F}\right)}&\\
&\Downarrow&\\
&L_{f}h + L_{g_1}u
+{1\over 2}{\rm Tr}\left\{\Sigma^{\T}g^{\T}{{\partial^2h}\over{\partial x^2}}g\Sigma\right\}
\geq -\alpha(h)\,,& 
\end{array}
\end{equation}
for system \eqref{eqn:su} with \eqref{eq-Sigma}, by feedback $u=u_0+\bar u(x,u_0)$ for a nominal contro law $u_0$, and call this condition the {\em noise-to state barrier function condition} (NSBFc). 

Heretofore, we have dealt with CBF and DSSf-CBFs. We say that a function $h$ is a {\em noise-to-state safety control barrier function} (NSSf-CBF) if, in addition to its usual conditions, it satisfies the implication
\begin{eqnarray}
L_{g_2}h=0 \quad\Rightarrow\quad \omega \geq 0 \,,
\end{eqnarray}
where
\begin{align}\label{eq-omega-NSSf}
\omega(x,u_0) =& L_{f+g_2 u_0}h(x) + \alpha(h)
\nonumber\\
&-{1\over 2}\left|g_1^{\T}{{\partial^2h}
\over{\partial x^2}}g_1\right|_{\cal F}\rho^{-1}(\max\left\{ 0,-h(x)\right\})
\end{align}
for a  class ${\cal K}$ $\rho: [0,+\infty) \rightarrow [0, -\inf h(\xi))$  and  $\alpha\in{\cal K}_h$.

\begin{theorem}\label{C3thm2.1s}
Under either the Sontag-type control  
\eqref{C3eq8.18}, \eqref{C3eq8.18kappa},
or the QP control 
\eqref{eq-QP-DSSf}, 
along with $\omega(x,u_0)$ defined in \eqref{eq-omega-NSSf}, 
  the system \eqref{eqn:su}  with \eqref{eq-Sigma} satisfies the NSBFc in \eqref{eq-NSBFc}. 
\end{theorem}

\noindent{\bf Proof.} 
For both control laws, a direct substitution yields $
{\cal L} h \geq -\alpha(h(x)) $ whenever $\min\left\{ 0, h(x)\right\}  \leq-\rho\left(\left|\Sigma\Sigma^{\T}\right|_{\cal F}\right)$. 
\hfill $\Box$

\begin{example}\label{C3exam8.1asNSSf}\em
To illustrate a design for NSSf, we return to Example \ref{ex-1+x2d}
but with the disturbance $d$ replaced by white noise of unknown covariance $\sigma(t)$, namely, to 
\begin{eqnarray}
{\rm d}{x} & = & u\, {\rm d}t+(1+x^2) \sigma(t)\, {\rm d}w \,.\label{C3exa8.1asNSSf}
\end{eqnarray}
To vary the design a bit but still keep it simple, we choose $h(x)=-x^3$ and $\alpha(h) = 3h$. 
Conducting the calculations with \eqref{eq-omega-NSSf}, with arbitrary $\rho\in{\cal K}_\infty$, we arrive at the QP safety filter
\begin{equation}\label{eq-QP-DSSf2ex+s}
\bar u_{\rm QP} = \min\left\{u_0, - (1+x^2)^2\rho^{-1}\left(\max\{0,|x|x\}\right)  - x\right\}\,.
\end{equation}
\hfill $\Box$
\end{example}

Next, we give a result on inverse optimal NSSf filter design. 

\begin{theorem}\label{C5thm:inop}
Consider the control law
\begin{align}
u=\ & u_0 + \bar u(x,u_0)\\
\bar u(x,u_0) =\ &
R_2^{-1}\left(L_{g_2}h\right)^{\T}
{{\ell\gamma_2\left(\left |L_{g_2}hR_2^{-{1/2}}\right |\right)}\over
{\left |L_{g_2}hR_2^{-{1/2}}\right |^2}},\label{C5eqn:clu}
\end{align}  
where $h(x)$ is a barrier function candidate,
$\gamma_1$ and $\gamma_2$ are class ${\cal K}_\infty$
functions whose derivatives are also class ${\cal K}_\infty$ functions, and $R_2 (x,u_0 )$ is a 
matrix-valued function such that $R_2(x,u_0)=R_2(x,u_0)^{\T}>0$. If the control law (\ref{C5eqn:clu}) 
makes the system 
\begin{eqnarray}
{\rm d} x=f(x)\, {\rm d} t+g_1(x)d\bar{w}+g_2(x)u\, {\rm d} t\label{C5eqn:ausys}
\end{eqnarray}  
satisfy the SBFc with respect to an NSSf-CBF candidate $h (x )$, where $\bar{w}$ is an independent $r$-dimensional stochastic
process with incremental covariance\index{incremental covariance}
\begin{eqnarray}
\bar{\Sigma}\bar{\Sigma}^{\T}=-2g_1^{\T}{{\partial^2h}\over{\partial x^2}}g_1
{{\ell\gamma_1\left(\left|g_1^{\T}{{\partial^2h}\over{\partial x^2}}g_1\right|_{\cal F}\right)}
\over{\left|g_1^{\T}{{\partial^2h}\over{\partial x^2}}g_1\right|^2_{\cal F}}},\label{C5eqn:ausigma}
\end{eqnarray}  
namely, if the condition
\begin{align}
&L_{f+g_2u_0}h-\ell\gamma_1\left(\left|g_1^{\T}{{\partial^2h}\over{\partial x^2}}g_1\right|_{\cal F}\right)
+\ell\gamma_2\left(\left |L_{g_2}hR_2^{-{1/2}}\right |\right) 
\nonumber\\
&\geq-\alpha(h)\,,
\end{align}
is satisfied, then the control law
\begin{align}
u=\ & u_0 + \bar u^*(x,u_0)\\
\bar u^*=\ &{\beta\over 2}R_2^{-1}\left(L_{g_2}h\right)^{\T}
{{\left(\gamma_2'\right)^{-1}\left(\left |L_{g_2}hR_2^{-{1/2}}\right |\right)}\over{\left |L_{g_2}hR_2^{-{1/2}}\right |}},\;\;\beta\geq 2\label{C5eqn:clu*}
\end{align}  
{\em maximizes} the cost functional
\begin{align}
J(u-u_0)=\ &\inf_{\Sigma\in{\cal D}}\left\{\lim_{t\rightarrow\infty}E\left[2\beta h(x(t))
\right.\right.\nonumber\\
&\left.\left.+\int_0^t\left(l(x,u_0)-\beta^2\gamma_2\left({2\over\beta}\left |R_2^{1/2}(u-u_0)\right |\right)\right.\right.\right.\nonumber\\
&\left.\left.\left.+\beta\lambda\gamma_1\left({{\left|\Sigma\Sigma^{\T}\right|_{\cal F}}\over{\lambda}}\right)\right)
d\tau\right]\right\},\label{C5eqn:opf}
\end{align}
where $\lambda\in(0,2]$ and
\begin{eqnarray}
l(x,u_0)\!\!&\!\!=\!\!&\!\! 
-2\beta \left[L_{f+u_0}h
-\ell\gamma_1\left(\left|g_1^{\T}{{\partial^2h}\over{\partial x^2}}g_1\right|_{\cal F}\right)
\right.\nonumber\\
&&\left.+\ell\gamma_2\left(\left|L_{g_2}hR_2^{-{1/2}}\right |\right)
\right]
\nonumber\\
\!\!&\!\!\!\!&\!\!-\beta \left(\beta -2\right)\ell\gamma_2\left(\left |L_{g_2}hR_2^{-{1/2}}\right |\right)
\nonumber\\
&&-\beta(2-\lambda)\ell\gamma_1\left(\left|g_1^{\T}{{\partial^2h}
\over{\partial x^2}}g_1\right|_{\cal F}\right)
\nonumber\\
&\leq & 2\beta\alpha(h)\,.
\end{eqnarray}
\end{theorem}

\noindent{\bf Proof.} 
According to Dynkin's formula\index{Dynkin's formula} and by  substituting 
$l (x,u_0 )$ into $J (u -u_0)$, we have
\begin{eqnarray}
&&J (u -u_0)
\nonumber\\
&=&\inf_{\Sigma\in{\cal D}}\left\{\lim_{t\rightarrow\infty}E\left[2\beta h(x(t))
\right.\right.\nonumber\\
&&\left.\left.+\int_0^t\left(l(x,u_0)-\beta^2\gamma_2\left({2\over\beta}\left |R_2^{1/2}(u-u_0)\right |\right)\right.\right.\right.\nonumber\\
&&\left.\left.\left.
+\beta\lambda\gamma_1\left({{\left|\Sigma\Sigma^{\T}\right|_{\cal F}}\over{\lambda}}\right)\right)
d\tau\right]\right\}\nonumber\\
&=&\inf_{\Sigma\in{\cal D}}\left\{\lim_{t\rightarrow\infty}E\left[2\beta h(x(0))
\right.\right.\nonumber\\
&&\left.\left.+\int_0^t\left(2\beta {\cal L}h\mid_{(\ref{eqn:su})}+l (x,u_0)
\right.\right.\right.\nonumber\\
&&\left.\left.\left.-\beta^2\gamma_2\left({2\over\beta}\left |R_2^{1/2}(u-u_0)\right |\right)
\right.\right.\right.\nonumber\\
&&\left.\left.\left.+\beta\lambda\gamma_1\left({{\left|\Sigma\Sigma^{\T}\right|_{\cal F}}\over{\lambda}}\right)
\right)d\tau\right]\right\}\nonumber\\
&=&\inf_{\Sigma\in{\cal D}}\left\{2\beta E\left\{V (x (0 ) )\right\}
\right.\nonumber\\
&&\left.+\lim_{t\rightarrow\infty}E\int_0^t\left[-\beta^2\gamma_2\left({2 \over \beta}
\left |R_2^{1/2}(u-u_0)\right |\right)
\right.\right.\nonumber\\
&&\left.\left.-\beta^2\ell\gamma_2\left(\left |L_{g_2}hR_2^{-{1/2}}\right |\right)
+2\beta L_{g_2}h(u-u_0)
\right.\right.\nonumber\\
&&\left.\left.+\beta\lambda\gamma_1\left({{\left|\Sigma\Sigma^{\T}\right|_{\cal F}}\over{\lambda}}\right)
+\beta\lambda\ell\gamma_1\left(\left|g_1^{\T}{{\partial^2h}\over{\partial x^2}}g_1\right|_{\cal F}\right)
\right.\right.\nonumber\\
&&\left.\left.+\beta{\rm Tr}\left\{\Sigma^{\T}g_1^{\T}
{{\partial^2h}\over{\partial x^2}}g_1\Sigma\right\}\right]d\tau\right\}.\label{C5eqn:ju}
\end{eqnarray}
Using Lemma \ref{lema.4} we have
\begin{align}
&-2\beta L_{g_2}h(u-u_0)\nonumber\\
&=\beta^2\left({2\over\beta}R_2^{1/2}(u-u_0)\right)^{\T}
\left(-R_2^{-1/2}\left(L_{g_2}h\right)^{\T}\right)\nonumber\\
&\leq \beta^2\gamma_2\left({2 \over \beta}\left |R_2^{1/2}(u-u_0)\right |\right)
+\beta^2\ell\gamma_2\left(\left |L_{g_2}hR_2^{-{1/2}}\right |\right)
\end{align}
and
\begin{align}
&\beta{\rm Tr}\left\{\Sigma^{\T}g_1^{\T}{{\partial^2h}\over{\partial x^2}}g_1\Sigma\right\}
\nonumber\\
&=\beta\left({\rm col}\left(\Sigma\Sigma^{\T}\right)\right)^{\T}
\left({\rm col}\left(g_1^{\T}{{\partial^2h}\over{\partial x^2}}g_1\right)\right)\nonumber\\
&\leq \beta\lambda\gamma_1\left({{\left|\Sigma\Sigma^{\T}\right|_{\cal F}}\over\lambda}\right)
+\beta\lambda\ell\gamma_1\left(\left|g_1^{\T}{{\partial^2h}\over{\partial x^2}}g_1\right|_{\cal F}\right)
\end{align}
and the equalities hold when \eqref{C5eqn:clu*} 
and
\begin{equation}
\left(\Sigma\Sigma^{\T}\right)^*=-\lambda(\gamma_1')^{-1}\left(\left|g_1^{\T}{{\partial^2h}
\over{\partial x^2}}g_1\right|_{\cal F}\right)
{{g_1^{\T}{{\partial^2h}\over{\partial x^2}}g_1}\over
{\left|g_1^{\T}{{\partial^2h}\over{\partial x^2}}g_1\right|_{\cal F}}}.\label{C5eqn:worst}
\end{equation}
So the ``worst case'' unknown covariance is given by (\ref{C5eqn:worst}),
the minimum of (\ref{C5eqn:ju}) is reached with $u=\bar u^*$, and
\begin{eqnarray}
\min_{u-u_0}J (u -u_0 )&=&2\beta E\left\{h (x (0 ) )\right\}.
\end{eqnarray}

\begin{remark}\label{C5remark:HJI}{\rm 
Similar to Remarks \ref{C3rem8.1} and \ref{C3rem8.1s}, even though not explicit in the statement of Theorem \ref{C5thm:inop}, $h (x )$ solves the following
family of {\it Hamilton-Jacobi-Isaacs} equations\index{Hamilton-Jacobi-Isaacs equations} parameterized by $\beta\in [2,\infty)$ 
and $\lambda\in(0,2]$:
\begin{align}
L_{f+u_0}V-{\lambda\over 2}\ell\gamma_1\left(\left|g_1^{\T}{{\partial^2h}\over{\partial x^2}}
g_1\right|_{\cal F}\right)
+{\beta \over 2}\ell\gamma_2\left(\left |L_{g_2}hR_2^{-{1/2}}\right |\right)\nonumber\\
+{l (x )\over{2\beta}}=0.\label{C5eqn:HJI}
\end{align}
This equation, which depends only on known quantities, helps explain why we are pursuing a
differential game\index{differential game} formulation for safe control design, with $\Sigma$ as a player. }\mbox{} \hfill $\Box$
\end{remark}

\begin{remark}\label{rem-attens}\em
Similar to Remark \ref{rem-atten}, we refer to the property
\begin{align}\label{C3eq8.14as}
\lim_{t\rightarrow\infty}E\left\{ h(x(t))+ \int_0^t  \left[\alpha(h(x)) 
+{\lambda\over 2}\gamma_1\left({{\left|\Sigma\Sigma^{\T}\right|_{\cal F}}\over{\lambda}}\right)\right] \, {\rm d} t\right\} \geq 0\,
\end{align}
as {\em integral noise-to-state safety} (iNSSf). 
\hfill $\Box$
\end{remark}

\section{Inverse Optimal Safety Filters for Adaptive Control, System Identification, and Extremum Seeking}
\label{sec-projection}

Most problems involving estimation of unknown parameters---be it in system identification, adaptive control, or extremum seeking---involve optimization. Safe sets of unknown parameters are not just about 
producing estimates that are within a set in which the unknown parameter is known to be. There is a more critical reason for keeping the parameters in a ``safe set'' in adaptive control---the safe set typically contains parameter values that correspond to the system model being controllable or stabilizable. Employing parameter estimates from outside of the safe set in indirect adaptive control results in an attempt of stabilizing an unstabilizable system, the result of which is the controller gains assuming infinite values. Hence, keeping parameters inside a safe set is critical in these domains of control theory. 

Keeping parameters inside a safe set is an add on to parameter estimator design. The typical problem of safety maintenance is simple. The parameter estimator has the form of a vector integrator
\begin{equation}\label{eq-update}
\dot{\hat\theta} = u
\end{equation}
where $\hat\theta \in \mathbb{R}^n$ is the parameter estimate and $u=u_0(\hat\theta,\xi,t)$ is the parameter estimator feedback, designed using gradient, least-squares, Lyapunov, passivity, or some other method, and possibly dependent on an additional state $\xi$, which may contain the states of the plant, an observer, and filters. The safety objective is formulated as keeping $\hat\theta$ inside the set $\left\{h(\hat\theta)\geq 0\right\}$, where $h$ has the usual properties of a CBF. 

The conventional safety filter in parameter estimation is {\em parameter projection}. Parameter projection assigns
\begin{equation}
u = u_0 + \bar u
\end{equation}
where $\bar u=\bar u_{\rm P}$ and 
\begin{equation}\label{eq-Proj}
\bar u_{\rm P} = \displaystyle{\left(\displaystyle{\partial h\over\partial\hat\theta}\right)^T \over \left| \displaystyle{\partial h\over\partial\hat\theta}\right|^2}
\left\{ \begin{array}{ll}
0, 
&  \alpha\left(h\left(\hat\theta\right)\right) > 0 \mbox{ or }
\displaystyle{\partial h\over\partial\hat\theta} u_0\geq 0 \vspace{3ex}\\
-\displaystyle{\partial h\over\partial\hat\theta} u_0, &
\alpha\left(h\left(\hat\theta\right)\right) = 0 \mbox{ and }
\displaystyle{\partial h\over\partial\hat\theta} u_0 <0\,
\end{array}\right.
\end{equation}
and $\alpha\in{\cal K}$ is arbitrary, typically taken as identity. 

Let us contrast this with the safety filter obtained using the QP approach  where $\bar u=\bar u_{\rm QP}$ and 
\begin{equation}\label{eq-ProjQP}
\bar u_{\rm QP}=\displaystyle{\left(\displaystyle{\partial h\over\partial\hat\theta}\right)^T \over \left| \displaystyle{\partial h\over\partial\hat\theta}\right|^2}
\max\left\{0, -\displaystyle{\partial h\over\partial\hat\theta} u_0 - \alpha\left(h\left(\hat\theta\right)\right) \right\}\,.
\end{equation}
Recall Remark \ref{rem-Lg2h=} on the notational convention of impossibility of division by ${\partial h\over\partial\hat\theta}=0$ in \eqref{eq-ProjQP}. 

The similarity between \eqref{eq-Proj} and \eqref{eq-ProjQP} is striking and not noted before in the literature, as  \eqref{eq-ProjQP} has not seen use in parameter estimation. While \eqref{eq-ProjQP} is continuous, \eqref{eq-Proj} interferes less with the nominal $u_0$ (it lets the trajectory come to the boundary and then glide tangentially if $u_0$ demands an exit from the safe set) but is discontinuous (at the boundary of the safe set). Intuitively, if one were to take $\alpha(r)$ outside of class ${\cal K}$, as a mapping such that $\alpha(0)=0$  but $\alpha(r) = +\infty$ for all $r>0$, one would obtain \eqref{eq-Proj} from  \eqref{eq-ProjQP}. One can approximate the projection operator quite closely by \eqref{eq-ProjQP} if one takes $\alpha(r) = {1\over\epsilon} r^\epsilon$ for sufficiently small positive $\epsilon$. 

Neither  \eqref{eq-Proj} nor \eqref{eq-ProjQP} have optimality properties but the following result, proven using Corollary \ref{C3thm8.1e}, holds for  \eqref{eq-ProjQP}. 

\begin{theorem}
The update law \eqref{eq-update} with 
\begin{equation}\label{C3eq8.5est-filter}
u = u_0 + \beta\bar u_{\rm QP}
\end{equation}
and \eqref{eq-ProjQP}, for any $\beta\geq 2$, {\em minimizes} 
\begin{align}\label{C3eq8.5est}
J(u-u_0) =& \lim_{t\rightarrow \infty}
\Bigg[-2\beta h\left(\hat\theta\right)
+\int_0^t \left( l\left(\hat\theta,u_0\right)
\right.\nonumber\\ &\left.\left.
+ {\left| \displaystyle{\partial h\over\partial\hat\theta}\right|^2\left|u-u_0\right|^2\over \max\left\{0, -\displaystyle{\partial h\over\partial\hat\theta} u_0 - \alpha\left(h\left(\hat\theta\right)\right) \right\}}
\right){\rm d}\tau \right]\,,
\end{align}
where
\begin{align}\label{eq-l-QP3}
l\left(\hat\theta,u_0\right) = & 2\beta \displaystyle{\partial h\over\partial\hat\theta} u_0+\beta^2  \max\left\{0, -\displaystyle{\partial h\over\partial\hat\theta} u_0 - \alpha\left(h\left(\hat\theta\right)\right) \right\}
\nonumber \\
\geq & - 2\beta \alpha\left(h\left(\hat\theta\right)\right)\,.
\end{align}
\end{theorem}

The cost functional \eqref{C3eq8.5est} clearly favors the update law $u$ staying close to the nominal design $u_0$. The two occurrences of a negative of $h(x)$ in \eqref{C3eq8.5est}, first in the ``terminal penalty'' before the integral and, second, in the lower bound on the state penalty $l(x,u_0)$ should be understood as measures of ``non-safety'' of the parameter estimator. By minimizing these non-safety measures, the safety filter \eqref{C3eq8.5est-filter} maximizes the estimator's safety.  

To summarize the update law, 
\eqref{eq-update}, \eqref{C3eq8.5est-filter}, \eqref{eq-ProjQP}, we get
\begin{equation}\label{eq-ProjQPupdate}
\dot{\hat\theta} = u_0 + \beta\displaystyle{\left(\displaystyle{\partial h\over\partial\hat\theta}\right)^T \over \left| \displaystyle{\partial h\over\partial\hat\theta}\right|^2}
\max\left\{0, -\displaystyle{\partial h\over\partial\hat\theta} u_0 - \alpha\left(h\left(\hat\theta\right)\right) \right\}\,
\end{equation}
for any $\beta\geq 2$. The ``soft projection'' in (E.5), (E.6) of  \cite{krstic1995nonlinear} is a special case of the update law \eqref{eq-ProjQPupdate}.

\section{Stabilization to the Safety Boundary}\label{sec-boundary}

In \cite{krstic2006nonovershooting}, stabilization to the safety boundary was pursued, under the name of ``non-overshooting'' control. In Section IV of that paper, a problem of disturbance-to-state stabilization was tackled, where convergence to the boundary was the goal but the achievement of regulation to a neighborhood of the boundary proportional to the unknown bound on the disturbance was guaranteed. 

In these results, the safe set was very special---it was ``half-space''---and in this situation, if a safety filter had been designed, it would have given $\bar u(x,u_0)=0$ since the nominal $u_0$ already ensures safety, in spite of the fact that it drives the trajectories to the boundary. 

In addition, the objective of $u_0$ in \cite{krstic2006nonovershooting} was also  specific---stabilization of a unique equilibrium on the boundary of the safe set, with that boundary being a hyperplane. 

In general, one can expect the boundary of the safe set to be more general than a hyperplane and also that the objective of $u_0$ be stabilization of the entire boundary, namely, regulation to the boundary (on which there may not even be an equilibrium) rather than to a unique equilibrium on the boundary. 

We pursue in this section a generalization of non-overshooting control from \cite{krstic2006nonovershooting} to systems with disturbances,
\begin{equation} \label{C3eq2.4non}
\begin{array}{rcl}
\dot x & = & f (x) + g_1 (x)d + g_2 (x) u\,.
\end{array}
\end{equation}
We don't employ any Lyapunov functions and we don't consider nominal $u_0$ whose task is equilibrium stabilization. Instead, the objective of $u_0$ is regulation of the barrier function $h(x(t))$ to zero, just as the objective of the safety filter $\bar u(x,u_0)$ shall be to prevent $h(x(t))$ from approaching zero too fast. 

To summarize, we consider a single barrier function $h(x)$ with two contradictory objectives but with two distinct input functions to be designed: $u_0(x)$ tasked with reducing $h(x)$ to zero and $\bar u(x,u_0)$ tasked with keeping $h(x)$ away from zero for all finite time. We approach these two simultaneous tasks with the following two functions:
\begin{eqnarray}\label{C3eq8.19nom-om0}
\omega_0(x) &=& L_{f+g_2 u_0(x)} h + \left| L_{g_1}h\right| \rho_0^{-1}(\max\{0, h(x)\})
\nonumber\\
&& +\alpha_0(h(x))
\\  \label{C3eq8.19nom-om}
\omega(x,u_0) &=& L_{f+g_2 u_0}h  - \left| L_{g_1}h\right| \rho^{-1}(\max\{0, -h(x)\})
\nonumber\\
&& +\alpha(h(x))\,,
\end{eqnarray}
where $\rho_0: [0,+\infty) \rightarrow [0, \sup h(\xi))$ and $\rho: [0,+\infty) \rightarrow [0, -\inf h(\xi))$ are in class ${\cal K}$ and $\alpha_0,\alpha\in{\cal K}_h$. 

We urge the reader to closely examine the similarities and differences between these two functions and then proceed. 

\begin{assumption}\label{ass-om0}
The feedback law $u_0(x)$ is input-to-output stabilizing for the system \eqref{C3eq2.4non} from the disturbance $d$ to the CBF $h(x)$ as an output, namely,
\begin{equation}
\omega_0(x) \leq 0\,,\quad \forall x\in\IR^n\,.
\end{equation}
\end{assumption}

\begin{assumption}\label{ass-om}
For a given nominal feedback law $u_0(x)$, the function $h(x)$ is a DSSf-CBF for system \eqref{C3eq2.4non}, namely,
\begin{equation}
L_{g_2}h(x) = 0 \quad\Rightarrow\quad \omega(x,u_0(x)) \geq 0\,,\quad \forall x\in\IR^n\,.
\end{equation}
\end{assumption}

\begin{assumption}\label{ass-alphas}
For all $x\in\IR^n$, 
\begin{align}
&\alpha(h(x)) \geq \alpha_0(h(x)) \nonumber\\
&+ \left| L_{g_1}h\right| \left[ \rho^{-1}(\max\{0, -h(x)\})+ \rho_0^{-1}(\max\{0, h(x)\})\right]
\end{align}
\end{assumption}

The meanings of Assumptions \ref{ass-om0} and \ref{ass-om} are obvious. Assumption \ref{ass-alphas} means that the system is such that its control can impart a sufficient margin of safety relative to the  nominal control $u_0(x)$ designed to drive the system to the safety boundary. The systems in \cite{krstic2006nonovershooting}, as well as their rather complicated backstepping designs of $u_0(x)$, satisfied all these assumptions. 

A generalization of non-overshooting control is given next. 

\begin{theorem}\label{thm-DSSf-sandwich}
Under Assumptions \ref{ass-om0}, \ref{ass-om}, and \ref{ass-alphas}, there exist functions $\beta\leq\beta_0$ of class ${\cal KL}_h$ such that
\begin{equation}\label{eq-QP-DSSfsfnon}
u = u_0(x) + (L_{g_2}h)^T
\displaystyle{\max\left\{0, -\omega(x,u_0(x))\right\}\over |L_{g_2}h|^2} \,,
\end{equation}
ensures the following for system \eqref{C3eq2.4non} for all $t\geq 0$: 
\begin{eqnarray} \label{C3eq2.5bQPa}
&\beta(h(x_0),t) - \rho\left(\sup_{0\leq \tau\leq t} |d(\tau)|\right)&\\[2mm]
&\leq h(x(t)) \leq & \nonumber\\[2mm]
&\beta_0(h(x_0),t) + \rho_0\left(\sup_{0\leq \tau\leq t} |d(\tau)|\right)\,.&
\label{C3eq2.5bQPb}
\end{eqnarray}
\end{theorem}

\noindent{\bf Proof.} Result \eqref{C3eq2.5bQPa} follows from Theorem \ref{thm-QP-DSSf}. 
To prove \eqref{C3eq2.5bQPb}, we substitute \eqref{eq-QP-DSSfsfnon} and \eqref{C3eq8.19nom-om} into \eqref{C3eq2.4non} and  get
\begin{eqnarray}
\dot h  &=& L_{f+g_2 u_0}h + L_{g_1} h d + L_{g_2} h \bar u_{\rm QP}
\nonumber\\
&=& -\alpha_0(h(x)) +\omega_0 +\max\left\{0, -\omega\right\}
\nonumber\\
&& -|L_{g_1}h|\rho^{-1}(\max\{0, h(x)\}) +L_{g_1} h d 
\nonumber\\
&\leq& -\alpha_0(h(x)) +\max\left\{\omega_0,\omega_0-\omega\right\}
\nonumber\\ &&
-|L_{g_1}h|\left[\rho_0^{-1}(\max\{0, h(x)\}) - |d|\right]\,. 
\end{eqnarray}
Since $\omega_0 - \omega = \alpha_0 - \alpha \leq 0$ and $\omega_0\leq 0$, we have that $\max\left\{\omega_0,\omega_0-\omega\right\}\leq 0$ and, hence,
\begin{equation}
\dot h \leq -\alpha_0(h(x)) -|L_{g_1}h|\left[\rho_0^{-1}(\max\{0, h(x)\}) - |d|\right]\,. 
\end{equation}
With an argument as, for example, in the proof of Theorem 2.2 in \cite{Krstic-Deng-book}, \eqref{C3eq2.5bQPb} follows. 
\hfill $\Box$

For the disturbance-free version of \eqref{C3eq2.4non}, namely, for $g_1=0$, Theorem \ref{thm-DSSf-sandwich} yields the following corollary.

\begin{corollary}\label{thm-DSSf-sandwich-d=0}
For a given feedback law $u_0(x)$ for system
\begin{equation}
\dot x = f(x ) + g(x) u\,, 
\end{equation}
let there exist functions $\alpha_0\leq \alpha$  in class ${\cal K}$ such that, for all $x\in\IR^n$, $\omega_0(x) := L_{f+g u_0(x)} h  +\alpha_0(h(x)) \leq 0$ and $\omega(x,u_0) := L_{f+g u_0}h   +\alpha(h(x))\geq 0$ whenever $L_{g}h=0$. Then there exist functions $\beta\leq \beta_0$  of class ${\cal KL}$ such that
\begin{equation}\label{eq-QP-DSSfsfnon-d=0}
u = u_0(x) + (L_{g}h)^T
\displaystyle{\max\left\{0, -\omega(x,u_0(x))\right\}\over |L_{g}h|^2} \,,
\end{equation}
ensures the following for all $t\geq 0$: 
\begin{eqnarray} 
\beta(h(x_0),t) \leq h(x(t)) \leq \beta_0(h(x_0),t) \,.
\label{C3eq2.5bQPd}
\end{eqnarray}
\end{corollary}

A similar non-overshooting design can be applied to the stochastic system \eqref{eqn:su}  with unknown covariance \eqref{eq-Sigma}. Inspired by \eqref{eq-omega-NSSf}, the functions in \eqref{C3eq8.19nom-om0} and \eqref{C3eq8.19nom-om} are modified, respectively, to 
\begin{align}\label{C3eq8.19nom-om0st}
\omega_0(x) =\ & L_{f+g_2 u_0(x)} h +\alpha_0(h(x))
\nonumber\\
&
+ {1\over 2}\left|g_1^{\T}{{\partial^2h}\over{\partial x^2}}g_1\right|_{\cal F} \rho_0^{-1}(\max\{0, h(x)\})
\\  \label{C3eq8.19nom-omst}
\omega(x,u_0) =\ & L_{f+g_2 u_0}h +\alpha(h(x)) 
\nonumber\\
&
- {1\over 2}\left|g_1^{\T}{{\partial^2h}\over{\partial x^2}}g_1\right|_{\cal F} \rho^{-1}(\max\{0, -h(x)\})
\,.
\end{align}
Denoting $\underline\rho(r) = \min\{\rho_0(r), \rho(r)\}$, it can be proven that feedback \eqref{eq-QP-DSSfsfnon} guarantees that, for all $x\in\IR^n$,
\begin{eqnarray} 
& 
\left|\Sigma\Sigma^{\T}\right|_{\cal F} \geq \underline\rho(|h(x)|)
& \nonumber\\
&\Rightarrow \ -\alpha(h(x)) \leq {\cal L}h(x) \leq -\alpha_0(h(x))\,.&\nonumber
\label{C3eq2.5bQPdnss}
\end{eqnarray}

Finally, for the covariance  $\Sigma(t)\Sigma(t)^{\T} \equiv I$, the stochastic non-overshooting-in-the-mean results in \cite{WuquanStochasticNonovershooting} are generalized as follows. 
Consider the  stochastic system \eqref{eqn:su} and, for a given feedback law $u_0(x)$, let there exist functions $\alpha_0\leq \alpha$  in class ${\cal K}$ such that, for all $x\in\IR^n$, $\omega_0(x) := L_{f+g_2 u_0(x)} h  +{1\over 2}{\rm Tr}\left\{g_1^{\T}{{\partial^2h}\over{\partial x^2}}
g_1\right\}+\alpha_0(h(x)) \leq 0$ and $\omega(x,u_0) := L_{f+g_2 u_0}h   +{1\over 2}{\rm Tr}\left\{g_1^{\T}{{\partial^2h}\over{\partial x^2}}
g_1\right\}+\alpha(h(x))\geq 0$ whenever $L_{g_2}h=0$. Then \eqref{eq-QP-DSSfsfnon} ensures that, for all $x\in\IR^n$,
\begin{eqnarray} 
-\alpha(h(x)) \leq {\cal L}h(x) \leq -\alpha_0(h(x))\,.
\label{C3eq2.5bQPd+}
\end{eqnarray}

In this section, $h$ played a twofold role of a barrier and Lyapunov function. For a general method for simultaneous (vector) Lyapunov-like functions for the same system, see \cite{6519314}.

\section{Conclusions}

For nonlinear systems affine in control, deterministic disturbance, or stochastic disturbance we introduced the appropriate notions of CBFs, designed safety-ensuring filters, and produced parametrized families of safety filters that have inverse optimality properties. Optimality is always such that the safety filter is rewarded for increasing safety and for keeping the input close to  nominal, while the disturbance is rewarded for decreasing safety and for not spending high energy.

Regarding the many variants on QP that we present, those that are pointwise optimal are not optimal over time, in the sense that they are faithful to the nominal control but are ``insufficiently safe'' to be optimal,   and vice versa, those safety filters that are inverse optimal (safe enough) are not pointwise optimal (not as `alert'). 

 Theorems \ref{C3thm8.1} and \ref{C5thm:inop}, as well as Example \ref{ex-1+x2d+}, show the potential benefit of stepping beyond the confines of the QP/min-norm design. 


To provide as complete a perspective on a range of problems and classes of systems, within the page limit, our technical exposition is concise and we leave illustration through simulation examples for future work, specialized to applications. 


\section*{Acknowledgments}

The author thanks Iasson Karafyllis, Xiangru Xu, and Andrew Clark for helpful feedback on a draft of the paper. 

\appendix
\section{Legendre-Fenchel Transform and Young's Inequality}


\begin{lemma}\label{lema.3}
If $\gamma$ and its derivative $\gamma^{'}$ are class 
${\cal K}_\infty$ , then the Legendre-Fenchel 
transform satisfies the following properties:
\begin{align}
(a) &\ \   \ell \gamma (r) = r (\gamma^{'})^{-1}(r) - \gamma\left((\gamma^{'})^{-1}(r)\right)= \int_0^r (\gamma^{'})^{-1}(s) ds \label{app1}\\
(b) &\ \ \ell\ell\gamma=\gamma\label{app2}\\
(c) &\ \ \ell \gamma \ \mbox{is a class ${\cal K}_\infty$ function}\\
(d) &\ \ \ell\gamma(\gamma^{'}(r))=r\gamma^{'}(r)-\gamma(r)\,.\label{app3}
\end{align}
\end{lemma}

\begin{lemma}\label{lema.4} 
{\bf (Young's inequality)} For 
any $x,y\in\mathbb{R}^n$, 
\begin{eqnarray}
x^{\T}y \leq \gamma(|x|)+\ell\gamma(|y|)\,,\label{app5}
\end{eqnarray}
and the equality is achieved if and only if
\begin{eqnarray}
y=\gamma^{'}(|x|)\frac{x}{|x|}\,, \ \mbox{that is,} \ \mbox{for} \ 
x=(\gamma^{'})^{-1}(|y|)\frac{y}{|y|}\,.
\end{eqnarray}
\end{lemma}

\bibliographystyle{IEEEtranS}
\bibliography{paper}

\begin{thebibliography}{10}
\providecommand{\url}[1]{#1}
\csname url@samestyle\endcsname
\providecommand{\newblock}{\relax}
\providecommand{\bibinfo}[2]{#2}
\providecommand{\BIBentrySTDinterwordspacing}{\spaceskip=0pt\relax}
\providecommand{\BIBentryALTinterwordstretchfactor}{4}
\providecommand{\BIBentryALTinterwordspacing}{\spaceskip=\fontdimen2\font plus
\BIBentryALTinterwordstretchfactor\fontdimen3\font minus
  \fontdimen4\font\relax}
\providecommand{\BIBforeignlanguage}[2]{{%
\expandafter\ifx\csname l@#1\endcsname\relax
\typeout{** WARNING: IEEEtranS.bst: No hyphenation pattern has been}%
\typeout{** loaded for the language `#1'. Using the pattern for}%
\typeout{** the default language instead.}%
\else
\language=\csname l@#1\endcsname
\fi
#2}}
\providecommand{\BIBdecl}{\relax}
\BIBdecl

\bibitem{DSCC}
I.~Abel, M.~Jankovi{\'c}, and M.~Krsti{\'c}, ``Constrained control of input
  delayed systems with partially compensated input delays,'' \emph{ASME Dynamic
  Systems and Controls Conference (DSCC)}, 2020.

\bibitem{9467052}
H.~Almubarak, N.~Sadegh, and E.~A. Theodorou, ``Safety embedded control of
  nonlinear systems via barrier states,'' \emph{IEEE Control Systems Letters},
  vol.~6, pp. 1328--1333, 2022.

\bibitem{almubarak2021hjb}
H.~Almubarak, E.~A. Theodorou, and N.~Sadegh, ``Hjb based optimal safe control
  using control barrier functions,'' 2021.

\bibitem{AmesCruiseControl}
A.~D. Ames, J.~W. Grizzle, and P.~Tabuada, ``Control barrier function based
  quadratic programs with application to adaptive cruise control,'' in
  \emph{IEEE Conference on Decision and Control}, 2014, pp. 6271--6278.

\bibitem{AmesAutomotive}
A.~D. Ames, X.~Xu, J.~W. Grizzle, and P.~Tabuada, ``Control barrier function
  based quadratic programs for safety critical systems,'' \emph{IEEE
  Transactions on Automatic Control}, vol.~62, pp. 3861--3876, 2017.

\bibitem{ANGELI1999209}
D.~Angeli and E.~D. Sontag, ``Forward completeness, unboundedness
  observability, and their lyapunov characterizations,'' \emph{Systems \&
  Control Letters}, vol.~38, no.~4, pp. 209--217, 1999.

\bibitem{basar-bernhard}
T.~Ba\c{s}ar and P.~Bernhard, \emph{$H^\infty$-Optimal Control and Related
  Minimax Design Problems: A Dynamic Game Approach}.\hskip 1em plus 0.5em minus
  0.4em\relax Birkhauser, 1998.

\bibitem{doi:10.1137/1.9781611971132}
T.~Ba\c{s}ar and G.~J. Olsder, \emph{Dynamic Noncooperative Game Theory{\em,
  2nd edition}}.\hskip 1em plus 0.5em minus 0.4em\relax Society for Industrial
  and Applied Mathematics, 1998.

\bibitem{8263977}
S.~Bansal, M.~Chen, S.~Herbert, and C.~J. Tomlin, ``Hamilton-jacobi
  reachability: A brief overview and recent advances,'' in \emph{2017 IEEE 56th
  Annual Conference on Decision and Control}, 2017, pp. 2242--2253.

\bibitem{breeden2021high}
J.~Breeden and D.~Panagou, ``High relative degree control barrier functions
  under input constraints,'' 2021.

\bibitem{9147721}
Y.~Chen, M.~Ahmadi, and A.~D. Ames, ``Optimal safe controller synthesis: A
  density function approach,'' in \emph{2020 American Control Conference
  (ACC)}, 2020, pp. 5407--5412.

\bibitem{choi2021robust}
J.~J. Choi, D.~Lee, K.~Sreenath, C.~J. Tomlin, and S.~L. Herbert, ``Robust
  control barrier-value functions for safety-critical control,'' 2021.

\bibitem{CLARK2021Stochastic}
A.~Clark, ``Control barrier functions for stochastic systems,''
  \emph{Automatica}, vol. 130, p. 109688, 2021.

\bibitem{9303896}
M.~H. Cohen and C.~Belta, ``Approximate optimal control for safety-critical
  systems with control barrier functions,'' in \emph{2020 59th IEEE Conference
  on Decision and Control (CDC)}, 2020, pp. 2062--2067.

\bibitem{940927}
H.~Deng, M.~Krstic, and R.~Williams, ``Stabilization of stochastic nonlinear
  systems driven by noise of unknown covariance,'' \emph{IEEE Transactions on
  Automatic Control}, vol.~46, no.~8, pp. 1237--1253, 2001.

\bibitem{DENG1997151}
H.~Deng and M.~Krstic, ``Stochastic nonlinear stabilization Ñ ii: Inverse
  optimality,'' \emph{Systems \& Control Letters}, vol.~32, pp. 151--159, 1997.

\bibitem{doi:10.1137/S0363012993258732}
R.~A. Freeman and P.~V. Kokotovic, ``Inverse optimality in robust
  stabilization,'' \emph{SIAM Journal on Control and Optimization}, vol.~34,
  pp. 1365--1391, 1996.

\bibitem{MagnusCortesNonsmooth}
P.~Glotfelter, J.~Cortés, and M.~Egerstedt, ``Nonsmooth barrier functions with
  applications to multi-robot systems,'' \emph{IEEE Control Systems Letters},
  vol.~1, no.~2, pp. 310--315, 2017.

\bibitem{HsuBipedal}
S.-C. Hsu, X.~Xu, and A.~D. Ames, ``Control barrier function based quadratic
  programs with application to bipedal robotic walking,'' in \emph{2015
  American Control Conference (ACC)}, 2015, pp. 4542--4548.

\bibitem{ITO200259}
H.~Ito and R.~A. Freeman, ``Uniting local and global controllers for uncertain
  nonlinear systems: beyond global inverse optimality,'' \emph{Systems \&
  Control Letters}, vol.~45, no.~1, pp. 59--79, 2002.

\bibitem{JankovicDelay}
M.~{Jankovi{\'c}}, ``Control barrier functions for constrained control of
  linear systems with input delay,'' in \emph{2018 American Control Conference
  (ACC)}, June 2018, pp. 3316--3321.

\bibitem{JANKOVIC2018359}
M.~Jankovic, ``Robust control barrier functions for constrained stabilization
  of nonlinear systems,'' \emph{Automatica}, vol.~96, pp. 359--367, 2018.

\bibitem{6519314}
I.~Karafyllis and Z.-P. Jiang, ``Global stabilization of nonlinear systems
  based on vector control lyapunov functions,'' \emph{IEEE Transactions on
  Automatic Control}, vol.~58, no.~10, pp. 2550--2562, 2013.

\bibitem{AmesISSCBF}
S.~{Kolathaya} and A.~D. {Ames}, ``Input-to-state safety with control barrier
  functions,'' \emph{IEEE Contr. Syst. Lett.}, vol.~3, pp. 108--113, 2019.

\bibitem{Krstic-Deng-book}
M.~Krsti\'{c} and H.~Deng, \emph{Stabilization of Nonlinear Uncertain
  Systems}.\hskip 1em plus 0.5em minus 0.4em\relax Springer, 2000.

\bibitem{661589}
M.~Krstic and Z.-H. Li, ``Inverse optimal design of input-to-state stabilizing
  nonlinear controllers,'' \emph{IEEE Transactions on Automatic Control},
  vol.~43, no.~3, pp. 336--350, 1998.

\bibitem{krstic2006nonovershooting}
M.~Krstic and M.~Bement, ``Nonovershooting control of strict-feedback nonlinear
  systems,'' \emph{IEEE Transactions on Automatic Control}, vol.~51, no.~12,
  pp. 1938--1943, 2006.

\bibitem{krstic1995nonlinear}
M.~Krstic, P.~V. Kokotovic, and I.~Kanellakopoulos, \emph{Nonlinear and
  adaptive control design}.\hskip 1em plus 0.5em minus 0.4em\relax John Wiley
  \& Sons, Inc., 1995.

\bibitem{WuquanStochasticNonovershooting}
W.~Li and M.~Krstic, ``Mean-nonovershooting control of stochastic nonlinear
  systems,'' \emph{IEEE Transactions on Automatic Control}, 2020.

\bibitem{LI19971459}
Z.-H. Li and M.~Krstic, ``Optimal design of adaptive tracking controllers for
  non-linear systems,'' \emph{Automatica}, vol.~33, pp. 1459--1473, 1997.

\bibitem{lyu2020smallgain}
Z.~Lyu, X.~Xu, and Y.~Hong, ``Small-gain theorem for safety verification of
  interconnected systems,'' 2020.

\bibitem{8814799}
M.~Maghenem and R.~G. Sanfelice, ``Characterization of safety and conditional
  invariance for nonlinear systems,'' in \emph{2019 American Control Conference
  (ACC)}, 2019, pp. 5039--5044.

\bibitem{TamasCovid}
T.~G. {Moln{\'a}r}, A.~W. {Singletary}, G.~{Orosz}, and A.~D. {Ames},
  ``Safety-critical control of compartmental epidemiological models with
  measurement delays,'' \emph{IEEE Contr. Sys. Lett.}, vol.~5, pp. 1537--1542,
  2021.

\bibitem{nguyen2016exponential}
Q.~Nguyen and K.~Sreenath, ``Exponential control barrier functions for
  enforcing high relative-degree safety-critical constraints,'' in \emph{2016
  American Control Conference (ACC)}.\hskip 1em plus 0.5em minus 0.4em\relax
  IEEE, 2016, pp. 322--328.

\bibitem{¿ksendal2010stochastic}
B.~{\O}ksendal, \emph{Stochastic Differential Equations: An Introduction with
  Applications}, ser. Universitext.\hskip 1em plus 0.5em minus 0.4em\relax
  Springer Berlin Heidelberg, 2010.

\bibitem{935055}
Z.~Pan, K.~Ezal, A.~Krener, and P.~Kokotovic, ``Backstepping design with local
  optimality matching,'' \emph{IEEE Transactions on Automatic Control},
  vol.~46, no.~7, pp. 1014--1027, 2001.

\bibitem{PrajnaMultiStateDelay}
S.~{Prajna} and A.~{Jadbabaie}, ``Methods for safety verification of time-delay
  systems,'' in \emph{Proceedings of the 44th IEEE Conference on Decision and
  Control}, 2005, pp. 4348--4353.

\bibitem{PrajnaStochastic}
S.~Prajna, A.~Jadbabaie, and G.~J. Pappas, ``A framework for worst-case and
  stochastic safety verification using barrier certificates,'' \emph{IEEE
  Transactions on Automatic Control}, vol.~52, no.~8, pp. 1415--1428, 2007.

\bibitem{JankovicDriverIntent}
Y.~Rahman, M.~Jankovic, and M.~Santillo, ``Driver intent prediction with
  barrier functions,'' in \emph{2021 Amer. Contr. Conf.}, 2021, pp. 224--230.

\bibitem{rawlings2017model}
J.~Rawlings, D.~Mayne, and M.~Diehl, \emph{Model Predictive Control: Theory,
  Computation, and Design}.\hskip 1em plus 0.5em minus 0.4em\relax Nob Hill
  Publishing, 2017.

\bibitem{SantilloMulti}
M.~Santillo and M.~Jankovic, ``Collision free navigation with interacting,
  non-communicating obstacles,'' in \emph{2021 American Control Conference
  (ACC)}, 2021, pp. 1637--1643.

\bibitem{CooganStochastic}
C.~Santoyo, M.~Dutreix, and S.~Coogan, ``A barrier function approach to
  finite-time stochastic system verification and control,'' \emph{Automatica},
  vol. 125, p. 109439, 2021.

\bibitem{sepulchre1997constructive}
R.~Sepulchre, M.~Jankovi{\'c}, and P.~Kokotovi{\'c}, \emph{Constructive
  Nonlinear Control}.\hskip 1em plus 0.5em minus 0.4em\relax Springer, 1997.

\bibitem{28018}
E.~Sontag, ``Smooth stabilization implies coprime factorization,'' \emph{IEEE
  Transactions on Automatic Control}, vol.~34, no.~4, pp. 435--443, 1989.

\bibitem{SONTAG1989117}
E.~D. Sontag, ``A ÔuniversalÕ construction of artstein's theorem on nonlinear
  stabilization,'' \emph{Syst. Contr. Lett.}, vol.~13, pp. 117--123, 1989.

\bibitem{10.1016/j.automatica.2008.11.017}
K.~P. Tee, S.~S. Ge, and E.~H. Tay, ``Barrier lyapunov functions for the
  control of output-constrained nonlinear systems,'' \emph{Automatica},
  vol.~45, no.~4, p. 918Ð927, apr 2009.

\bibitem{WangMagnusMultiRobot}
L.~Wang, A.~D. Ames, and M.~Egerstedt, ``Safety barrier certificates for
  collisions-free multirobot systems,'' \emph{IEEE Transactions on Robotics},
  vol.~33, no.~3, pp. 661--674, 2017.

\bibitem{Wieland}
P.~Wieland and F.~Allg{\"{o}}wer, ``Constructive safety using control barrier
  functions,'' \emph{IFAC Proceedings Volumes}, vol.~40, no.~12, pp. 462 --
  467, 2007, 7th IFAC Symposium on Nonlinear Control Systems.

\bibitem{WuSreenathFirstCBFhighRelDeg}
G.~Wu and K.~Sreenath, ``Safety-critical and constrained geometric control
  synthesis using control lyapunov and control barrier functions for systems
  evolving on manifolds,'' in \emph{2015 American Control Conference (ACC)},
  2015, pp. 2038--2044.

\bibitem{xiao2019control}
W.~Xiao and C.~Belta, ``Control barrier functions for systems with high
  relative degree,'' in \emph{2019 IEEE 58th Conference on Decision and Control
  (CDC)}.\hskip 1em plus 0.5em minus 0.4em\relax IEEE, 2019, pp. 474--479.

\bibitem{XuRobustness}
X.~Xu, P.~Tabuada, J.~W. Grizzle, and A.~D. Ames, ``Robustness of control
  barrier functions for safety critical control.'' \emph{IFAC-PapersOnLine},
  vol.~48, no.~27, pp. 54 -- 61, 2015.

\bibitem{XU2018195}
X.~Xu, ``Constrained control of input-output linearizable systems using control
  sharing barrier functions,'' \emph{Automatica}, vol.~87, pp. 195--201, 2018.

\bibitem{XuAmesLaneKeeping}
X.~Xu, J.~W. Grizzle, P.~Tabuada, and A.~D. Ames, ``Correctness guarantees for
  the composition of lane keeping and adaptive cruise control,'' \emph{IEEE
  Transactions on Automation Science and Engineering}, vol.~15, no.~3, pp.
  1216--1229, 2018.

\bibitem{YIN2020104736}
H.~Yin, A.~Packard, M.~Arcak, and P.~Seiler, ``Reachability analysis using
  dissipation inequalities for uncertain nonlinear systems,'' \emph{Systems \&
  Control Letters}, vol. 142, p. 104736, 2020.

\bibitem{9126836}
H.~Yin, P.~Seiler, and M.~Arcak, ``Backward reachability using integral
  quadratic constraints for uncertain nonlinear systems,'' \emph{IEEE Control
  Systems Letters}, vol.~5, no.~2, pp. 707--712, 2021.

\end{thebibliography}


\begin{IEEEbiography}
{Miroslav Krstic}
(S'92-M'95-SM'99-F'02) 
 is Distinguished Professor, Alspach endowed chair, founding director of the  Center for Control Systems and Dynamics, and Senior Associate Vice Chancellor for Research at UC San Diego. 
 He is Fellow of IEEE, IFAC, ASME, SIAM, AAAS, IET (UK), and AIAA (Assoc. Fellow), as well as foreign member of the Serbian Academy of Sciences and Arts and of the Academy of Engineering of Serbia. He has received the Bellman Award, SIAM Reid Prize, ASME Oldenburger Medal, Nyquist Lecture Prize, Paynter Award, Ragazzini  Award, IFAC Nonlinear Control Systems Award, Chestnut Award, CSS Distinguished Member Award, the PECASE, NSF Career, and ONR YI awards, the Schuck (Õ96 and Õ19) and Axelby award, and the first UCSD Research Award given to an engineer. 
 He is EiC of Systems \& Control Letters and has been senior editor in Automatica, IEEE Trans. Automatic Control, and of two Springer book series. 
 Krstic has coauthored sixteen books on adaptive, nonlinear, and stochastic control, extremum seeking, control of PDE systems including turbulent flows, and control of delay systems.

\end{IEEEbiography}
\vfill
\end{document}